\documentclass[structabstract]{aa}
\usepackage{txfonts}
\usepackage{graphicx}
\usepackage{natbib}
\usepackage{nicefrac}
\usepackage{pgfplots}
\usepackage{booktabs}
\usepackage{multirow}
\usepackage{ulem}
\bibpunct{(}{)}{;}{a}{}{,}

\begin{document}

\title{Binary white dwarfs in the halo of the Milky Way} 

\author{Pim van Oirschot\inst{1}\thanks{\email{P.vanOirschot@astro.ru.nl}}
  \and Gijs Nelemans\inst{1,2} \and Silvia Toonen\inst{1,3} \and Onno Pols\inst{1} \and \\
  Anthony~G.~A. Brown\inst{3} \and Amina Helmi\inst{4} \and Simon Portegies Zwart\inst{3}} 

\institute{Department of Astrophysics/IMAPP, Radboud University Nijmegen,
  P.O. Box 9010, 6500 GL Nijmegen, The Netherlands 
  \and Institute for Astronomy, KU Leuven, 
  Celestijnenlaan 200D, 3001 Leuven, Belgium
  \and Leiden Observatory, Leiden University, 
  P.O. Box 9513, 2300 RA Leiden, The Netherlands
  \and Kapteyn Astronomical Institute, University of Groningen, 
  P.O. Box 800, 9700 AV, Groningen, The Netherlands} 

\date{Received 13 May 2014 / Accepted 30 June 2014}

\abstract {} {We study single and binary white dwarfs in the inner halo of the Milky Way
in order to learn more about the conditions under which the population
of halo stars was born, such as the initial mass function (IMF), the star formation history, 
or the binary fraction.} 
{ We simulate the evolution of low-metallicity halo stars at distances up to $\sim 3$ kpc
using the binary population synthesis code SeBa.
We use two different white dwarf cooling models to 
predict the present-day luminosities of halo
white dwarfs. We determine the white dwarf luminosity functions (WDLFs) for 
eight different halo models and compare these with the observed halo WDLF of white dwarfs in the SuperCOSMOS Sky Survey.
Furthermore, we predict the properties of binary white dwarfs in the halo and determine 
the number of halo white dwarfs that is expected to be observed with the Gaia
satellite.
} {By comparing the WDLFs, we find that a standard IMF matches the observations more accurately than
a top-heavy one, 
but the difference with a bottom-heavy IMF is small. 
A burst of star formation 13 Gyr ago fits slightly better than 
a star formation burst 10 Gyr ago 
and also slightly better than continuous star formation $10 - 13$ Gyrs ago.
Gaia will be the first instument 
to constrain the bright end of the field halo WDLF, where contributions from binary WDs are considerable. 
Many of these will have He cores, of which a handful have atypical surface gravities ($\log g<6$) and reach luminosities $\log (L/L_\odot) > 0$ 
in our standard model for WD cooling. These so called pre-WDs, if observed, can help us to constrain 
white dwarf cooling models and might teach us something about the fraction of halo stars that reside in binaries.
} {}

\keywords{Galaxy: halo -- Stars: luminosity function, mass function 
-- white dwarfs -- binaries: close} 
\titlerunning{Binary white dwarfs in the halo of the Milky Way}
\authorrunning{P. van Oirschot et al.}
\maketitle 

\section{Introduction}

The Galactic halo is the oldest component of our Galaxy,
containing metal-poor stars with high velocity dispersion.
It contains a small percent of the total stellar mass
of the Galaxy. Many questions about the formation of the halo 
and the Milky Way's oldest stars are still to be answered,
such as What is its star formation history (SFH)?, 
What is the initial mass function (IMF)?, 
and What fraction of halo stars resides in binaries?. In this paper, we will investigate
all three of these questions by studying the population of halo
white dwarfs with a population synthesis approach.

White dwarfs (WDs) are an increasingly important tool used to study Galactic populations.
Because they are the end product of low and intermediate mass stars, WDs
are interesting objects of study for age determinations
\citep[eg.][]{Hansen:2007,Bedin:2009}.
Since we have entered the era of large sky surveys, a huge amount of high-quality
observational data of these stars is now or will soon become available.
Physically, WDs are rather well understood, and they have been used 
as cosmic chronometers to study our Galaxy, as well as open and globular
clusters, for more than two decades (\citealt{Winget:1987}; see reviews by \citealt{Fontaine:2001}
and \citealt{Althaus:2010}, for example).
Since in general, halo WDs are cool and faint, we confine ourselves to studying the ones in the 
solar neighbourhood. 

The halo WD luminosity function (WDLF) was first derived by \citet{Liebert:1989},
based on six observed WDs with tangential velocities $v_\mathrm{t}$ exceeding 250 km s$^{-1}$.
The most recent estimate is based on observations of 93 WDs 
with $v_\mathrm{t} > 200$ km s$^{-1}$ in the SuperCOSMOS Sky Survey \citep[][hereafter RH11]{Rowell:2011}.
Theoretical halo WDLFs have been determined by, amongst others, \citet{Adams:1996,Isern:1998,Camacho:2007}.
Predictions for Gaia's performances on WDs have been made by \citet{Torres:2005}.
For a recent paper on this topic, see \citet{Carrasco:2014}.
However, the effect of binary stars has never been studied in great detail. 
Furthermore, different initial parameters, stellar evolution codes, and 
WD cooling models were used in most of these papers.

For different assumptions about the IMF, SFH, and binary fraction,
as well as for two different WD cooling models, we determine the WDLF 
and compare it with the observed halo WDLF in RH11.
We derive both its shape and its normalization 
from an independent mass density of low-mass halo stars. We include not only
single stars, but also focus on the contribution from WDs in binaries and WDs
that are the result of a binary merger. 
Furthermore, we predict the properties 
of the population of binary WDs in the halo for a standard model
and derive the number of halo WDs that can be detected
by the Gaia satellite.

The setup of this paper is as follows: in section~\ref{models} we explain our methods,
in section~\ref{results} we discuss our results, and our conclusions can be found in section~\ref{conclusions}.
In the concluding section we try to answer the question: 
What can Gaia observations of halo WDs teach us about the IMF, SFH, and binary fraction in the halo?

\vspace{1cm}

\section{Model ingredients} \label{models}

We aim to derive the WDLF from first principles, i.e. not normalizing it to the observed
WDLF, but using an independent estimate of the local stellar halo mass density to deduce a WDLF. 
A very important ingredient
of our model is therefore the relation between this local density $\rho_0$, the stellar halo mass
in the solar neighbourhood (the region that we simulate) and the IMF. In the next subsection,
the expected number of halo stars is derived for three different IMFs. More details on this 
calculation can be found in the two appendices of this paper.

\subsection{Initial Mass Functions} \label{IMFs}
As a standard assumption, the IMF $\phi(m)$ can be written as a power law
\begin{equation}
\phi(m) \equiv \frac{\mathrm{d}N}{\mathrm{d}m} \propto m^{-(\gamma + 1)} \label{dNdm}
\end{equation}
with $N$ being the number of stars formed in the mass range $m, m+\mathrm{d}m$ and $\gamma$ the IMF slope. 
We assume $\phi(m)$ to be independent of Galactic age or metallicity.
Unless specified otherwise, $N$ here represents the number of stars in the case that all stars are single 
(a binary fraction of 0). In section~\ref{2.2} we explain how these numbers change 
with a nonzero binary fraction.

In a classical paper, \citet{Salpeter:1955} estimated $\gamma = 1.35$, 
and the corresponding IMF is nowadays referred to as a Salpeter IMF. Although not our standard model,
one of the IMFs that we investigate in this paper is a Salpeter IMF for the whole
mass range of stars ($0.1-100$ M$_\odot$). It is nowadays generally believed
that the IMF flattens below 1.0 M$_\odot$, so this can be considered
a bottom-heavy IMF. 

In our standard model we split the IMF up into three power laws, following \citet{Kroupa:1993}:
\begin{equation}
\phi(m) \propto 
\left\{
    \begin{array}{ll}
\nicefrac{35}{19} \ m^{-1.3} &\qquad \mathrm{if} \ 0.1 \le m < 0.5, \\ 
\quad m^{-2.2} &\qquad \mathrm{if} \ 0.5 \le m < 1.0, \\ 
\quad m^{-2.7} &\qquad \mathrm{if} \ 1.0 \le m < 100. \label{Kroupa}
\end{array} \right.
\end{equation}

The thrid IMF that we investigate in this paper is the top-heavy IMF suggested 
by \citet{Suda:2013}.
These authors argued that the IMF for stars with [Fe/H] $<-2$ is lognormal
\begin{equation}
\phi(m) \propto \frac{1}{m} \ \exp\left[-\frac{\log_{10}^2 (m/\mu)}{2\sigma^2} \right] \label{eq:03}
\end{equation}
with median mass $\mu = 10$ and dispersion $\sigma = 0.4$.
Originally, this IMF was proposed by \citet{Komiya:2007} for stars with [Fe/H] $<-2.5$ 
to explain the observed features of carbon enhanced metal poor stars,
therefore we refer to it as the Komiya IMF.
The higher metallicity stars would be formed according to a Salpeter IMF.
Following the metallicity distribution function (MDF) of a two-component halo
model \citep{An:2013}, 24\% of the zero-age main-sequence (ZAMS) stars 
with masses between 0.65 and 0.75 M$_\odot$ are formed according to a Komiya IMF.
Therefore, when normalizing the WDLF properly, we expect more signatures from
high-mass WDs (which cool fast and are thus faint) when choosing this IMF.

In order to determine the actual number of stars in a population $N$, one has to integrate $\phi(m)$,
thereby setting the integration boundaries and the normalization constant.
For example, integrating equation~(\ref{dNdm}) with normalization constant $A$ yields
\begin{equation}
N = \int_{m_\mathrm{low}}^{m_\mathrm{high}} A \ m^{-(\gamma + 1)} \mathrm{d}m. \label{N}
\end{equation}
Hereafter, $m_\mathrm{low} = 0.1$ and $m_\mathrm{high} = 100$
for single stars and binary primaries.
The value of $A$ can be determined from an observed mass or number density of stars.
We use the estimated local stellar halo mass density $\rho_0 = 1.5 \cdot 10^{-4} \ \mathrm{M}_\odot \ \mathrm{pc}^{-3}$, 
based on observations of 16 halo stars in the mass range $0.1 \le m < 0.8$ by \citet{Fuchs:1998}.
These authors derived $\rho_0 = 1 \cdot 10^{-4} \ \mathrm{M}_\odot \ \mathrm{pc}^{-3}$ as a firm lower limit.
For a discussion on the correctness of this value compared to for example the lower estimate 
$\rho_0 = 6.4 \cdot 10^{-5} \ \mathrm{M}_\odot \ \mathrm{pc}^{-3}$ \citep{Gould:1998},
see \citet{Digby:2003} and \citet{Helmi:2008}.
Since 0.8 M$_\odot$ is roughly the mass below which all stars can be considered unevolved,
and 0.1 M$_\odot$ is our assumed lower mass boundary of all stars that are formed, this mass density
is directly related to the total mass in unevolved stars $M_\mathrm{unev}$ in our simulation box.
Our top-heavy IMF has two normalization constants, one for the very metal-poor stars and one for the 
higher metallicity stars. The normalization constants are derived in Appendix~\ref{Ap:B}.

\begin{figure*}
	\centering
		\includegraphics[height=0.24\textheight]{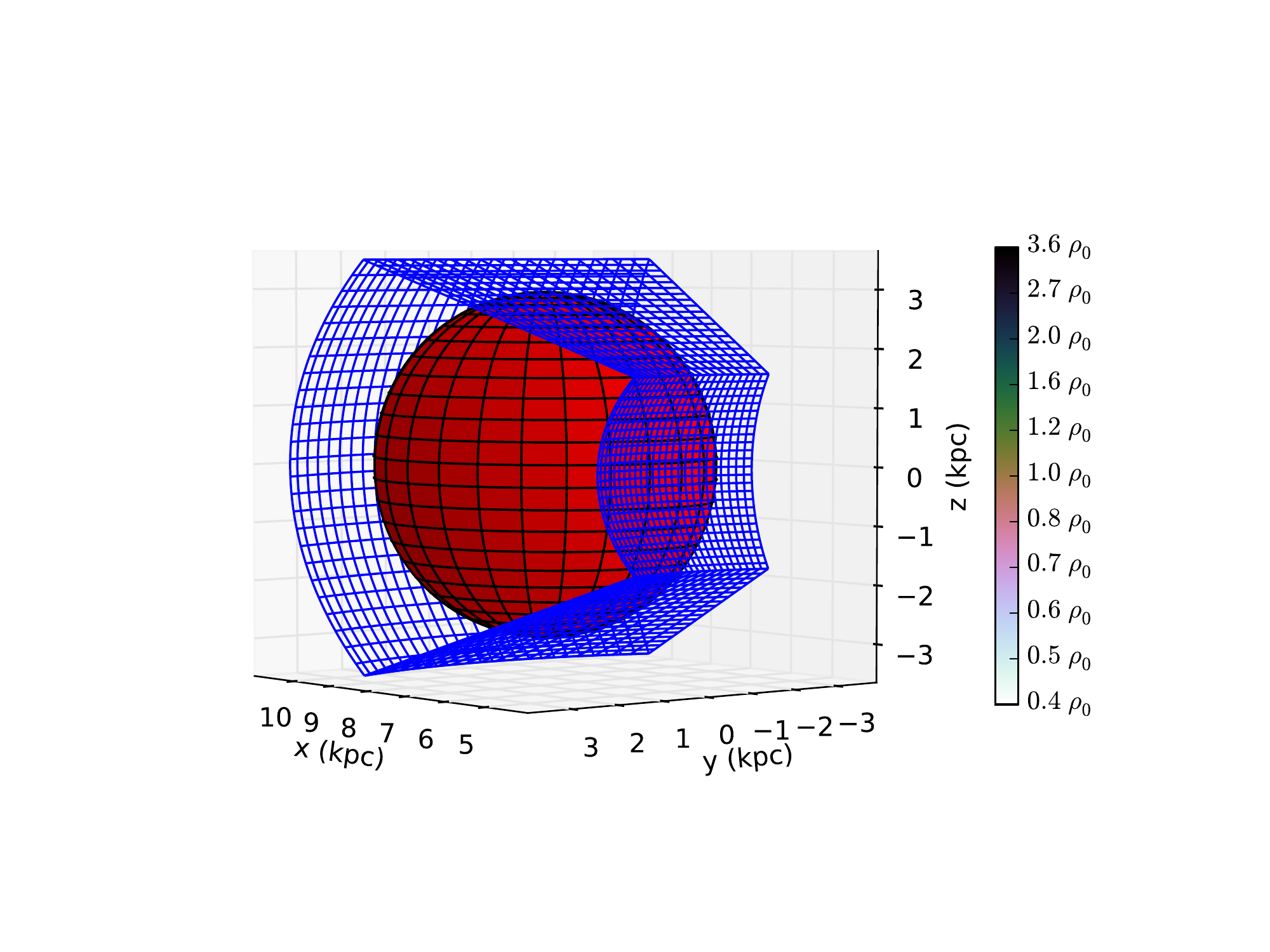}
		\includegraphics[height=0.24\textheight]{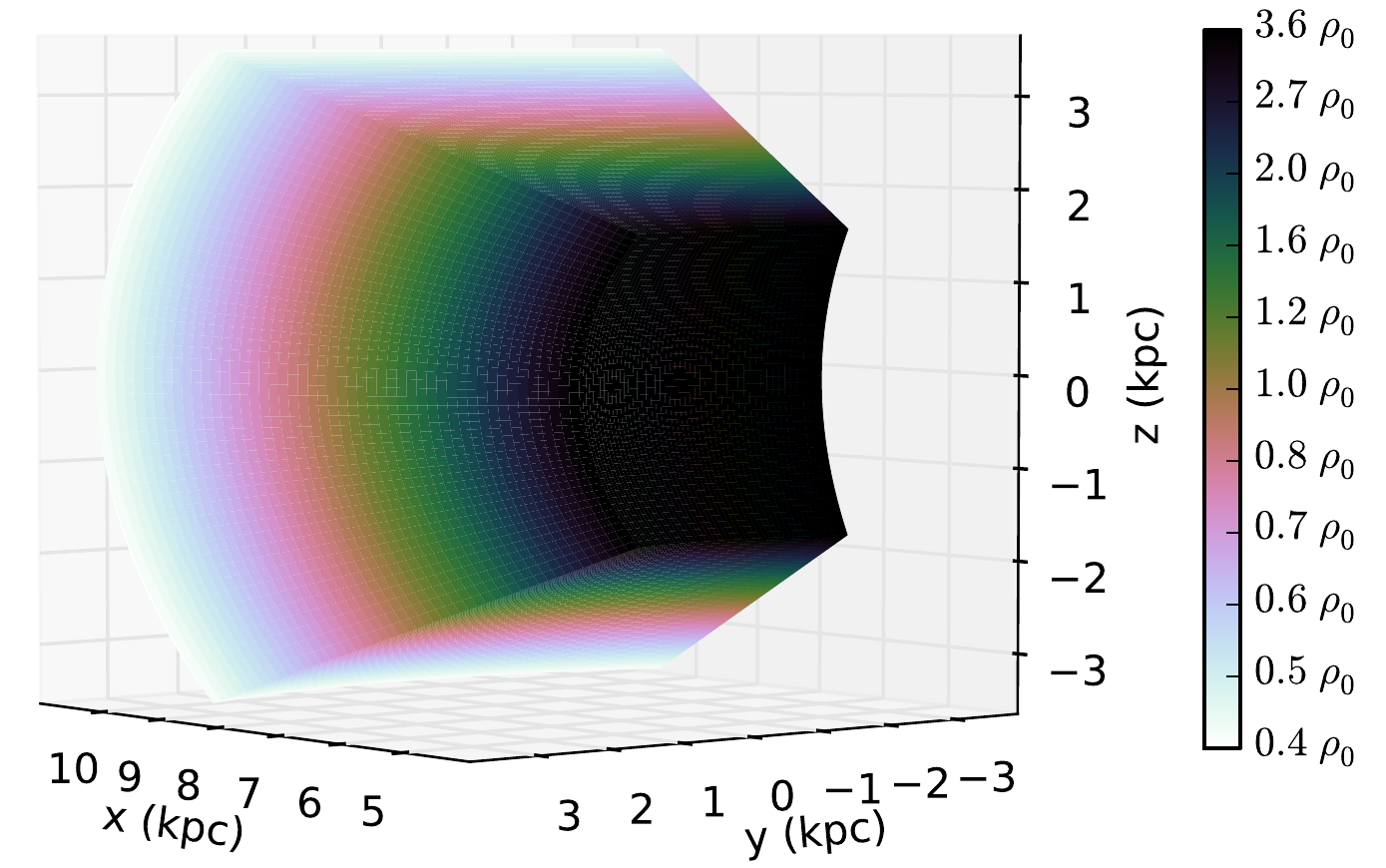}
		\caption{Simulation box containing a sphere with radius $\xi =2.95$ kpc centred at the position of the Sun $(x,y,z)_\odot = (8.0,0,0)$ kpc,
		$n = -2.8$ and $q = 0.64$ (left panel) and a density map of the simulation box (right panel).}
\label{01}
\end{figure*}

Our simulation box represents the stellar halo in the solar neighbourhood,
which we parameterize in a principal axis cartesian coordinate system as \citep{Helmi:2008}
\begin{equation}
\rho(x,y,z) = \frac{\rho_0}{r_0^n}\left(x^2 + y^2 + \frac{z^2}{q^2}\right)^{\nicefrac{n}{2}}, \label{eq.1}
\end{equation}
with $r_0$ the distance from the Sun to the Galactic centre,
$q$ the minor-to-major axis ratio and $n$ the power law exponent
of the density profile. Throughout this paper, an oblate stellar halo ($q <1$) is assumed.
A sphere with radius $\xi<r_0$ around the Sun defines the minimum width and height 
of our simulation box. We show in section~\ref{sec:7} that $\xi = 2.95$~kpc
is sufficient for our study of halo WDs.
Furthermore, we choose $r_0 = 8.0$ kpc
\citep[][an average of 16 literature measurements]{Moni-Bidin:2012},
$n=-2.8$ and $q = 0.64$ \citep{Juric:2008}.
The simulated area with these values of $r_0$, $\xi$, $n$ and $q$ is shown in Figure~\ref{01}.
We note that although $n=-2.8$ and $q = 0.64$ are the formal best-fit parameters of \citet{Juric:2008}, 
one should keep in mind the ranges $-3 \leq n \leq -2.5$ and $0.5 \leq q \leq 0.8$ as their fit results.
Substituting the above-mentioned value of $\rho_0$ into equation~(\ref{eq.1})
and integrating over the volume of our simulation box, we find 
$M_\mathrm{unev} = 3.6 \cdot 10^7 \ \mathrm{M}_\odot$ (see Appendix~\ref{Ap:A}).

A crucial part of the normalization is the mass function of low-mass stars.
The above-mentioned three IMFs predict drastically different numbers of stars
in the range of masses $0.1 \le m < 0.8$, which may or may not be in agreement with the
observed sample from which the local halo mass density is determined \citep{Fuchs:1998}.
Since the mass function cannot be determined indisputably from this sample of 16 stars,
we assume for each individual modelled IMF that it holds also in the low-mass regime.
However, the resulting number of expected halo WDs depends strongly on the normalization
of these low-mass stars (as we will see in section~\ref{sec:6}), so we also investigate the effect
of different slopes of the IMF at $0.1 \le m<0.8$.

Since the mass in low-mass stars is fixed by our normalization, the
number of evolved stars and thus of WDs depends on the ratio
of evolved stars to unevolved stars. The flatter the slope of the IMF for
unevolved stars, the fewer unevolved stars there are, i.e. the higher
this ratio\footnote{This statement also holds for the combined Salpeter+Komiya IMF,
but not for the Komiya IMF by itself.}.
Most studies of low-mass stars suggest that the slope of this part 
of the IMF flattens \citep[eg.][$\gamma_\mathrm{unev} \approx 0$]{Bonnell:2007}. 
This is why we take the Kroupa mass function as our standard.
Furthermore, we calculate the number of evolved stars,
which have initial masses $0.8 \le m < 100$, for a flat ($\gamma_\mathrm{unev} = -1$) IMF, 
yielding a robust upper limit on the number of evolved stars, $N_\mathrm{ev,upper}$ (see Appendix~\ref{Ap:B}).
In this way we derive a range of possible values for the number of evolved stars in our simulation box,
between $N_\mathrm{ev}$ (where $\gamma_\mathrm{unev}$ is set consistently by the IMF, also
in the low-mass regime) and $N_\mathrm{ev,upper}$ (derived by setting $\gamma_\mathrm{unev} = -1$).
These numbers for the three different assumptions about the IMF are given in 
the first three columns of Table~\ref{table:01}, assuming all stars are single.
In section~\ref{2.2} we give a discription of the last three columns of Table~\ref{table:01}.

\begin{table}
\caption{Number of stars in our simulation box as a function of the IMF.}
\begin{tabular}{ccccccc}
\toprule 
\multirow{2}{*}{IMF} &  $\underline{N_\mathrm{unev}}$ & $\underline{N_\mathrm{ev}}$ & $\underline{N_\mathrm{ev,upper}}$ & $\underline{N_{\mathrm{wd}(0)}}$ & $\underline{N_{\mathrm{wd}(0.5)}}$& $\underline{N_{\mathrm{wd}(1)}}$\\
&$10^7$&$10^7$&$10^7$&$10^7$&$10^7$&$10^7$\\
\midrule
 Kroupa &	   $12$  		& $1.9$			& $5.9$ 	& $1.7$		& $1.1$		& $0.70$	\\
\midrule
 Salpeter &    $17$  		& $1.1$			& $6.7$ 	& $0.97$	& $0.63$	& $0.41$	\\
\midrule
 Top-heavy & 	$16$  		& $80$			& $330$ 	& $25.4$	& $21.2$	& $18.3$	\\
 
 $-$ Komiya 	& $0.24$  	& $79$			& $326$ 	& $24.5$	& $20.6$	& $17.9$	\\
 
 $-$ Salpeter	& $16$  	& $1.0$			& $4.3$ 	& $0.93$	& $0.61$	& $0.39$	\\
\bottomrule
\end{tabular}
\tablefoot{\small For three different IMFs are indicated: 
in the first three columns, the number of stars in our simulation box ($\times 10^7$) 
for a binary fraction of 0 (all stars are single).
The border mass below which all stars are considerd to be unevolved is $0.8 \ \mathrm{M}_\odot$.
The numbers $N_\mathrm{ev,upper}$ come from taking $\gamma_\mathrm{unev} = -1$.
The resulting number of WDs with three different
assumptions on the binary fraction (0, 0.5, or 1) are listed in the last three columns,
with our standard assumptions on the SFH and WD cooling. 
For the top-heavy IMF, the first of the three lines lists the sum of the contributions from 
the Komiya and the Salpeter IMF, which are given in the second and third line.}
\label{table:01}
\end{table}

\subsection{Population synthesis}\label{2.2}
The evolution of the halo stars is calculated with the binary 
population synthesis code SeBa \citep{Portegies-Zwart:1996,Toonen:2012,Toonen:2013}.
In SeBa, stars are generated with a Monte-Carlo method, using the following distributions:
\begin{itemize}
\item Binary primaries are drawn from the same IMF as single stars
(see section~\ref{IMFs}).
\item Flat mass ratio distribution between 0 and 1, thus for secondaries $m_\mathrm{low} = 0$
and $m_\mathrm{high} = m_\mathrm{primary}$.
\item Initial separation: flat in $\log a$ (\"Opik's law) between 1 R$_\odot$ and $10^6$ R$_\odot$ \citep{Abt:1983},
provided that the stars do not fill their Roche lobe.
\item Initial eccentricity: chosen from the thermal distribution $\Xi(e) = 2e$
between 0 and 1 as proposed by \citet{Heggie:1975} and \citet{Duquennoy:1991}.
\end{itemize}

All simulated halo stars have metallicity $\mathrm{Z} = 10^{-3}$ (0.05 Z$_\odot$), unless 
a top-heavy IMF is assumed. In that case, all stars that are born following a
Komiya IMF are generated with metallicity $\mathrm{Z} = 2 \cdot 10^{-4}$ (0.01 Z$_\odot$).
The common-envelope (CE) presciption of the standard model in SeBa \citep[$\gamma \alpha$,][]{Toonen:2012} 
is used. With SeBa, we calculate the stellar evolution up to the point where the stars become WDs,
neutron stars, or black holes, as well as the evolution of the binary systems
until the end time of the simulation. For the WD cooling a separate code is 
used (see section~\ref{sec:3}).

Having determined the total stellar mass in the simulated area, 
we still need to make an assumption on the binary fraction in order to arrive at the total
number of stars in our simulation box. 
Because we assume a flat mass ratio distribution, the mass of the secondary is 
on average half the mass of the primary. The total number of binary systems in our simulation box
if all stars are in binaries is therefore 1.5 times less than the total number of ZAMS stars 
if the binary fraction is zero. 
As a standard assumption we adopt a binary fraction of 0.5. 
This means that there are as many binary systems as there are single stars, thus that
two out of every three stars are in a binary system. The total number of single stars
(which is the same as the total number of binary systems) in this case can be found by
dividing the numbers in the first three columns of Table~\ref{table:01} by 2.5.

The resulting number of WDs in our simulation box is listed in the last three columns
of Table~\ref{table:01} for three different values for the binary fraction (0, 0.5, or 1)
and standard assumptions about the SFH and WD cooling (see the next subsections).
These numbers are quite sensitive to the assumed binary fraction, 
especially for a Kroupa or Salpeter IMF,
because most of the binary primaries are unevolved stars in this case.
This means that even more secondaries are unevolved stars, which do not become WDs
within the age of the Galaxy, if the binary fraction is larger. Thus the total number
of WDs is smaller if the binary fraction is larger. For a top-heavy IMF this effect is
obviously less significant.

\begin{figure*}
\centering	
	\includegraphics[width=0.49\textwidth]{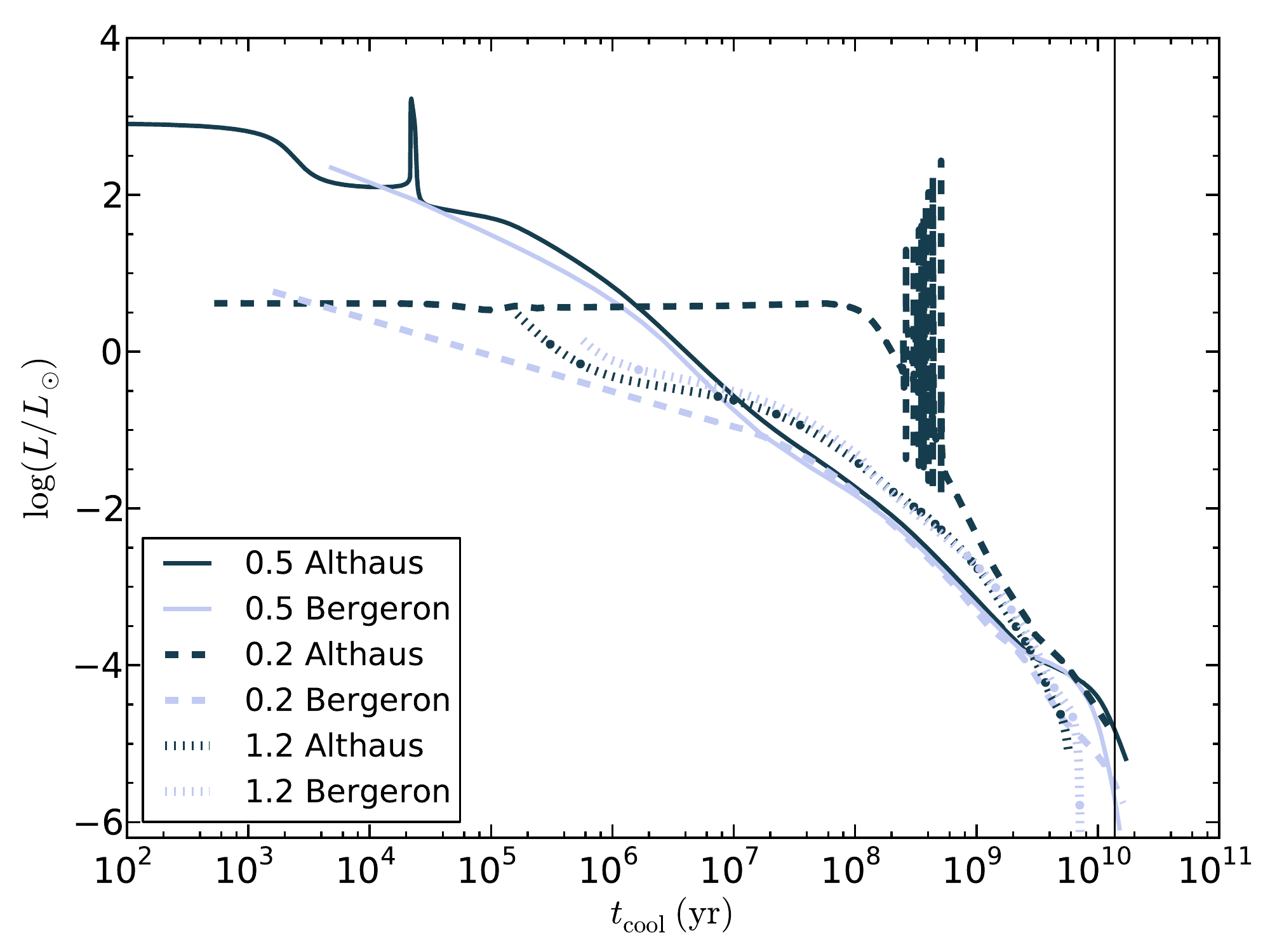}
	\includegraphics[width=0.49\textwidth]{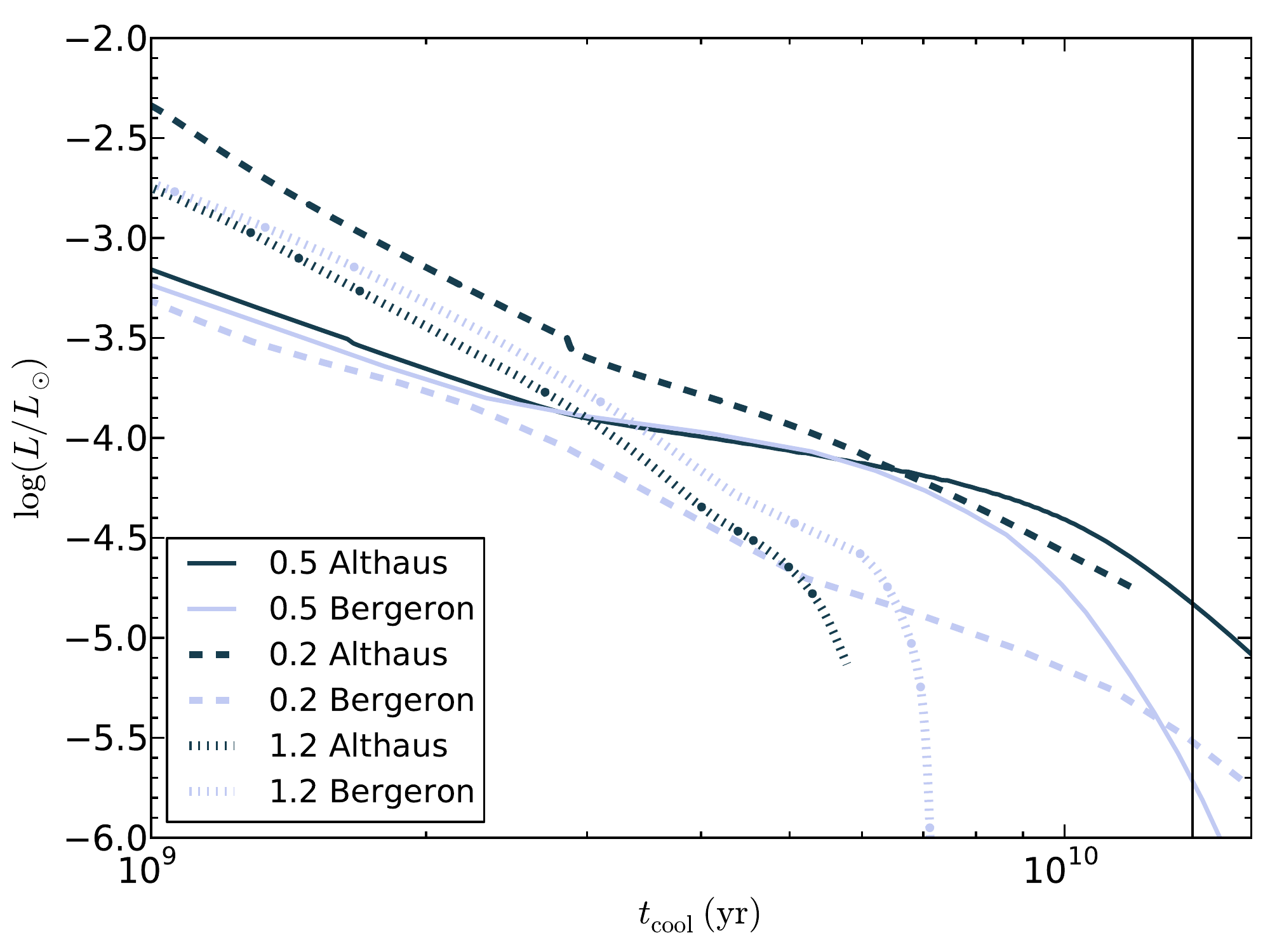}
	\caption{Comparison of cooling tracks of 0.2 M$_\odot$ (dashed lines), 0.5 M$_\odot$ (solid  lines) 
		and 1.2 M$_\odot$ (dotted lines) WDs, between the models from the Althaus group (dark blue lines) 
		and the Bergeron group (light blue lines).
		The 3 lines that represent cooling tracks from the Althaus models correspond to 3 WDs with different
		core compositions: He for the 0.2 M$_\odot$ WD, CO for the 0.5 M$_\odot$ WD and ONe for the 1.2 M$_\odot$ WD, 
		whereas the 3 lines from the Bergeron models all correspond to WDs with a CO core.
		The age of the universe is indicated with a vertical thin black line.}
		\label{02}
\end{figure*}

In our simulation we distinguish between carbon-oxygen core (CO) WDs, 
helium core (He) WDs and oxygen-neon core (ONe) WDs.
He WDs must have undergone episodes of mass loss in close
binary systems in order to be formed within a Hubble time.
They are thus only found in binary systems, or as a result of
two components of a binary system that merged.
In Table~\ref{table:02}, the mass ranges in which these three types
of WDs occur in our simulation are listed.
These mass ranges are partly overlapping, due to the effect of mass transfer
in close binary systems.
We note that they are dependent on the initial to final mass relation (IFMR),
and therefore using a different population synthesis code may affect these
results.
See \citet{Toonen:2014} for a comparison between four population synthesis codes.

\begin{table}
\caption{Minimum and maximum masses of the various types of WDs after 13 Gyr in our simulations.}
\begin{tabular}{lcccccc}
\toprule 
& \multicolumn{2}{c}{He} & \multicolumn{2}{c}{CO} & \multicolumn{2}{c}{ONe} \\ 
\cmidrule(l){2-3} \cmidrule(l){4-5} \cmidrule(l){6-7} 
& min & max & min & max & min & max \\
\midrule
single WD 		&  ---  &  ---  & 0.537 & 1.18 & 1.18 & 1.38 \\
double WDs   & 0.140 & 0.496 & 0.330 & 1.38 & 1.10 & 1.38 \\
merger product  	& 0.290 & 0.502 & 0.405 & 1.38 & 1.15 & 1.38 \\
\bottomrule
\end{tabular}
\tablefoot{\small A long dash (---) indicates that the particular combination does not occur. }
\label{table:02}
\end{table}

\subsection{Star Formation Histories}

We make three different assumptions about the star formation history of the halo,
based on observational indications that the vast majority
of halo stars are old \citep{Unavane:1996,Kalirai:2012}:
\begin{itemize}
\item[(a)] one burst of star formation 13 Gyr ago (our standard); 
\item[(b)] continous star formation from 13 until 10 Gyr ago, no star formation afterwards;
\item[(c)] one burst of star formation 10 Gyr ago; 
\end{itemize}
where a burst is assumed to last 250 Myr.
After SeBa is run, all simulated halo stars have the same age, i.e. 
all stars are evolved for 10 Gyr or for 13 Gyr.
To account for the SFH of the halo, we therefore shorten their lifetime 
by an amount of time randomly chosen between 0 and 250 Myr in case of assumptions (a) and (c) 
or with a number between 0 and 3 Gyr in case of assumption (b).
If, by doing so, the lifetime of the star is reduced below the time it takes that star
to become a WD, it is removed from our sample of WDs.

\subsection{White dwarf cooling models}\label{sec:3}

To determine the temperature, surface gravity and luminosity of a WD with a
certain mass and cooling time, we use the cooling tracks published by \cite{Althaus:2013} for He WDs 
and those from \cite{Renedo:2010} for CO WDs. 
For ONe WDs, we use the cooling tracks from \citet{Althaus:2007},
both to determine their luminosities and temperatures, and their colours and magnitudes.
We will refer to this set of cooling models as the Althaus models (our standard for WD cooling).
The metallicities of the He and CO WDs in the Althaus models are assumed to be 
$\mathrm{Z} = 0.01$, that of the ONe WDs $\mathrm{Z} =  0.02$.
Colours and magnitudes for CO WDs come from \citet{Althaus:2012}, whereas colour
tables of He WDs with high-metallicity progenitors ($\mathrm{Z} = 0.03$) \citep{Althaus:2009}
were used to determine the colours and magnitudes of He WDs.
In all these cooling tracks and colour tables, the WDs have a higher metallicity than the ones 
in our simulation box ($\mathrm{Z} = 0.001$). However, from the cooling tracks that were available for
different metallicities \citep{Althaus:2009,Panei:2007} we conclude that the effect of 
metallicity on WD cooling is smaller than other effects, such as the core composition (He or CO)
of the WD, at least for large cooling times.

Alternatively, we also use the WD cooling tracks that are published on the website
\texttt{www.astro.umontreal.ca/} $\sim$\texttt{bergeron/CoolingModels/}
\citep{Bergeron:2011,Holberg:2006,Kowalski:2006,Tremblay:2011}, hereafter
called the Bergeron models. The main difference between these two sets of cooling models
is that in the Althaus models the evolution prior to WD formation is taken into account 
to arrive at a WD with a certain core composition, whereas in the Bergeron models
the ad hoc assumption is made that all WDs have a CO core.

\begin{figure*}
  \centering 
  \resizebox{\hsize}{!}{\includegraphics{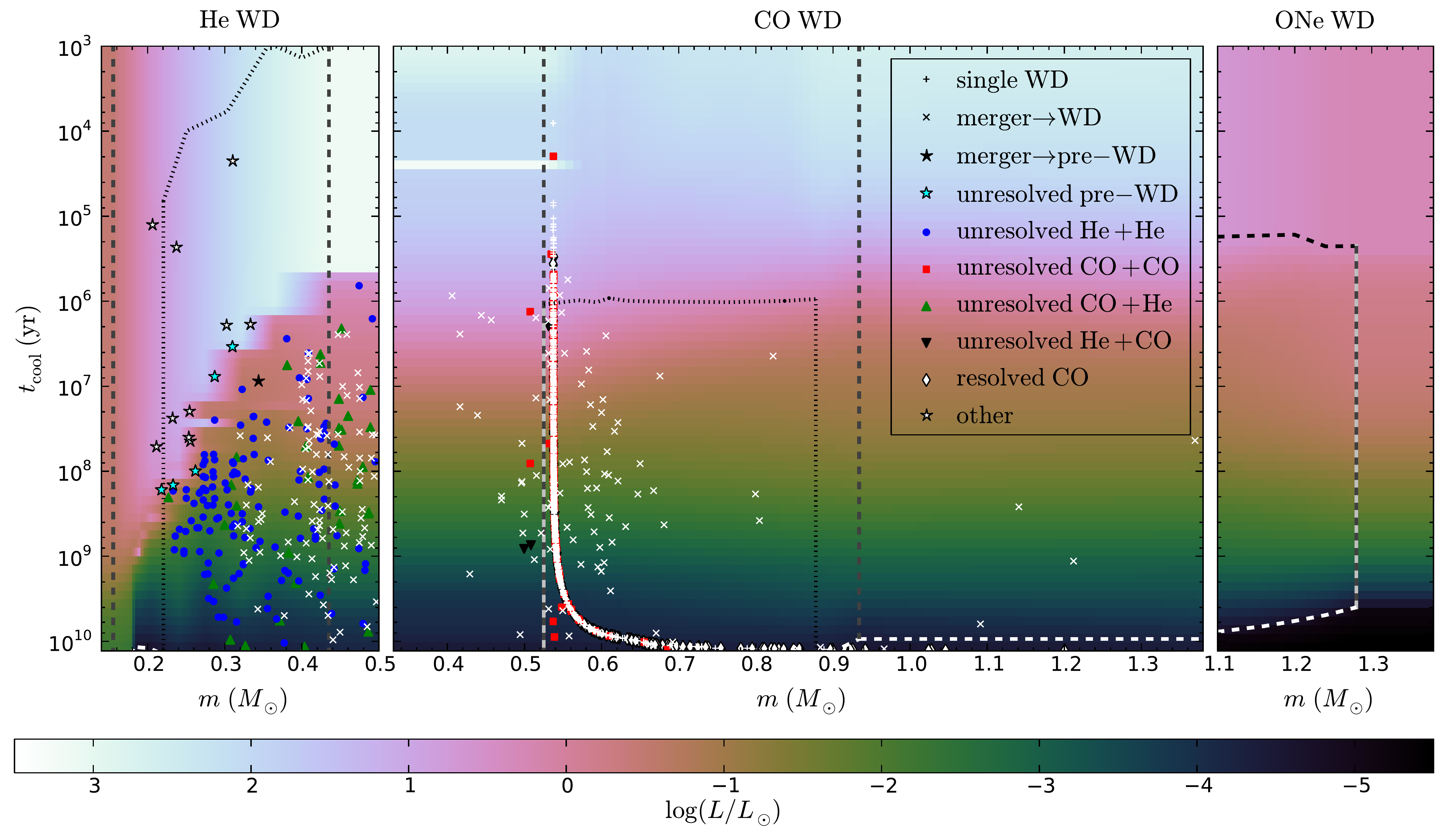}} 
  \caption{From left to right: constructed grids of He, CO and ONe WD cooling in the Althaus models.		
		The width of the panels corresponds to the maximum possible mass range per panel
		in our simulation, see Table~\ref{table:02}. Vertical (horizontal) dashed lines indicate masses 
		(cooling times) beyond which luminosities are obtained from extrapolation of the used cooling tracks. 
		Dotted lines indicate the boundaries beyond which $M_V$ and $M_I$ magnitudes are obtained by extrapolation.
		For ONe WDs these regions completely overlap, for CO WDs the lower mass boundary partly overlaps,
		which is indicated by the light-grey-dark-grey dashed line. 
		Scattered points indicate halo WDs that are observable with Gaia in our standard model.
		An explanation of the symbols is given section~\ref{sec:7}.}
  \label{03}
\end{figure*}

In order to compare the two cooling models, we take the low-mass end of the CO WDs in the Bergeron models 
as the ``He core'' WDs in their models, and the high-mass end as their ``ONe core'' WDs,
since non-CO core WD cooling models from Bergeron et al. were unfortunately not available
in the literature.
A comparison between these cooling models is given in Figure~\ref{02}.
For details about the CNO flashes, which are very prominent on the cooling branch of 
the He WD in the left panel of Figure~\ref{02}, we refer to \citet{Althaus:2013}.
Because the global specific heat of the He WDs is larger in the Althaus models, 
at a given cooling time the luminosity of such a low-mass WD is higher
than that of a WD with the same mass in the Bergeron models (compare the two dashed lines in Figure~\ref{02}).
Similarly, the global specific heat of ONe WDs is smaller in the Althaus models,
resulting in lower luminosity WDs at a given cooling time compared to the high-mass
WDs in the Bergeron models (the dotted lines in Figure~\ref{02}).

The explanation of the difference between the two solid lines in Figure~\ref{02} is
a bit more complicated, since in this case the WD core composition is the same in both models.
Whether the prior evolution of the WD is or is not taken into account,
will affect the onset of crystallization and the magnitude of the energy released 
by CO phase separation, a process that affects the cooling times below 
$\log (L/L_\odot) \approx -4$. This could account for part of the difference in the cooling times
between both models for the CO WDs. In addition,
at low luminosities, the WD evolution is sensitive to the treatment of the outer
boundary conditions and the equation of state at low densities \citep{Althaus:2012}.

Having chosen a set of cooling tracks, we want to determine the present-day luminosites 
for all the WDs in our model. 
Per WD type (He, CO, or ONe), typically only ten cooling tracks are available; these are interpolated
and extrapolated over mass and cooling time to cover the whole parameter space that is
sampled by our population synthesis code.
The interpolation is linear, both in mass and in cooling time. For WDs that are more massive
than the most massive WD of which a cooling track is available in the literature,
we assume the cooling to be the same as for the most massive that is available.
Similarly, the cooling track of the least massive WD is taken for WDs with a lower mass.
In the Althaus models, this ``extrapolation in mass'' is done for He WDs with $m<0.155$ or $m>0.435$, 
for CO WDs with $m<0.505$\footnote{The cooling track
corresponding to the 0.5~M$_\odot$ CO WD in the Althaus models that is plotted in Figure~\ref{02} 
is thus assumed to be equal to that of an 0.505~M$_\odot$ WD in these models.} or $m>0.934$, and for ONe WDs with $m>1.28$. 
At the faint end of the cooling track, for CO WDs with $m>0.878$ and $t_\mathrm{cool} \gtrsim 10^{10}$ years 
and ONe WDs with $t_\mathrm{cool} \gtrsim 6 \cdot 10^9$ years ($\pm 2 \cdot 10^9$ years, depending on the WD mass), 
the luminosity is extrapolated using \citet{Mestel:1952} cooling. 
If the cooling tracks that we use in this study do not 
give a value for the luminosity of the WD at birth, we keep the luminosity constant at 
the value corresponding to the first given cooling time (for ONe WDs, this is $\sim 10^5$ years).
This yields lower limits to the luminosities of very young ONe WDs in the Althaus models and 
for all types of very young WDs in the Bergeron models.

In this way we construct three grids of WD cooling (one for each WD type; He, CO, or ONe),
which are shown in the three different panels of Figure~\ref{03}. In this catalogue,
the luminosity of every star in our simulation box can be found.
The mass, type and cooling time of every WD in our simulation box was matched to the nearest catalogue
grid point using the K3Match software \citep{Schellart:2013}.
The dashed lines in Figure~\ref{03} indicate the boundaries beyond which extrapolation was done as described above.
Dotted lines similarly indicate the region beyond which the $M_V$ and $M_I$ magnitudes were determined
using extrapolation. The luminosity and colour regions that are covered 
by the available cooling tracks in the literature overlap for ONe WDs, 
and so does the $m=0.505$ boundary for CO WDs. This is indicated by the light grey dashes
in between the dark grey vertical lines at the overlapping points.
The scattered points in Figure~\ref{03} visualize the position of the halo WDs that are
observable with Gaia according to our standard stellar halo model. 
They will be discussed in section~\ref{sec:7}.

\subsection{Preparation of the WDLF}

In this paper, we distinguish between spatially resolved and unresolved binaries.
For each binary system, from the assigned distance and orbital separation
a separation on the sky can be determined. 
Assuming thus two stars in a binary should be separated by at least 0.1 - 0.2 arcsec
in order to be spatially resolved by Gaia \citep{Arenou:2005},
we assign all binaries with a separation larger than or equal to 0.3 arcseconds
to the group of resolved binaries, those with a smaller separation 
to the group of unresolved binaries. 
Unresolved binaries are included as a single WD in the 
WDLF with a luminosity equal to the
sum of the luminosites of the individual WDs in the binary.

To obtain a WDLF that can be compared with the observed one by RH11,
we transform the luminosities of the WDs in our simulation box to 
bolometric magnitudes (using $M_{\mathrm{bol},\odot} = 4.75$).
We divide the total magnitude range into 2 bins per magnitude,
ending up with bins such as $M_{\mathrm{bol},\odot} = [3.0,3.5]$ and $[3.5,4.0]$, etc.
The total number of stars per bin is then divided by the effective volume of our simulation box 
($V_\mathrm{eff} = M_\mathrm{unev}/\rho_0$) to arrive at $N \ \mathrm{pc}^{-3} \ M_\mathrm{bol}^{-1}$.

There are also observational selection effects that need to be taken into account.
Because RH11 only included halo WDs with tangential velocities 
$v_\mathrm{t} > 200$ km s$^{-1}$, we reduce the number of WDs in each 
luminosity bin by a factor $P(v_\mathrm{t} > 200)$, which represents
the probability that the tangential velocity of a halo star exceeds 200 km s$^{-1}$.
The tangential velocities of halo stars \citep{Chiba:2000} along the line of sight
to one of the SuperCOSMOS survey fields is shown in Figure~16 in RH11.
From this figure, we estimate that $P(v_\mathrm{t} > 200)$ lies
between 0.4 and 0.5, therefore we take 0.45. In our results section we will
show the effect of choosing $P(v_\mathrm{t} > 200) = 0.4$ or 0.5 instead.

\subsection{Gaia magnitudes and extinction} \label{Gaia}
The light from distant stars gets absorbed and reddened by interstellar dust.
Following \citet{Toonen:2013}, we assume that the dust follows the distribution
\begin{equation}
P(z) \propto \mathrm{sech}^2(z/z_h),
\end{equation}
where $z_h$ is the scale height of the Galactic dust (assumed to be 120 pc) and $z$ 
the cartesian coordinate in the $z$-direction.
The interstellar extinction $A_V$ between the Milky Way and a distant Galaxy in the $V$-band
is assumed to be the extinction between us and a star at a distance $d=\infty$ \citep{Sandage:1972},
for Galactic latitude $b = \arcsin (z/d)$:
\begin{eqnarray}
A_V = \left\{
    \begin{array}{ll}
      0.165 (\tan 50^\circ-\tan b)\csc b &\quad \mathrm{if} \ |b| < 50^\circ\\
      0 &\quad \mathrm{if} \ |b| \geq 50^\circ
    \end{array} \right\}
    \equiv A_V(\infty). \nonumber
\end{eqnarray}
The fraction of the extinction between us and a star at a distance $d$
with Galactic latitude $b \ne 0$ and this extinction is then
\begin{equation}
\frac{A_V(d)}{A_V(\infty)} =  \frac{\int_0^{d\sin b} P(z) \mathrm{d}z}{\int_0^\infty P(z) \mathrm{d}z} = \mathrm{tanh} \left(\frac{d\sin b}{z_h}\right). \label{Avd}
\end{equation}
The stars in our simulation box are distributed according to the density profile given by equation~(\ref{eq.1}),
from which the distance to these stars is determined as
\begin{equation}
d = \sqrt{(r_0-x)^2 + y^2 + z^2}.
\end{equation}
Equation~(\ref{Avd}) is used to calculate the apparent magnitude $V$ of a star at distance $d$
and Galactic latitude $b \ne 0$:
\begin{equation}
V = M_V + 5\left(\log_{10}(d)-1\right) + A_V(\infty) \ \mathrm{tanh} \left(\frac{d\sin b}{z_h}\right).
\end{equation}
Because $A_I(d) = 0.6 \ A_V(d)$ \citep{Schlegel:1998}, we similarly calculate
$I$ from $M_I$ and equation~(\ref{Avd}),
after which Gaia magnitudes are calculated using \citep{Jordi:2010}
\begin{equation}
G-V = a_1 + a_2 (V-I) + a_3 (V-I)^2 + a_4 (V-I)^3 
\end{equation}
with $a_1 = - 0.0257$, $a_2 =  - 0.0924$, $a_3 = - 0.1623$ and $a_4 = 0.0090$.
We expect Gaia to detect all WDs with $G<20$ \citep{Brown:2013}.

\section{Results} \label{results}

We start our results section with an analysis of the theoretically determined WDLF
in our standard halo model and compare it with the observationally determined one
by RH11. In the second part of this section, we will compare the WDLFs predicted 
by models that were introduced in section~\ref{models} and discuss our findings.
In section~\ref{sec:7}, we examine the halo WD population in more detail, again for our standard model.
We derive the number of halo WDs that will be detectable by the Gaia 
satellite, and also predict properties of the whole population of (binary) WDs in the halo.

\subsection{Standard model WDLF: theory vs. data}\label{sec:6}

\begin{figure*}
\centering	
	\resizebox{\hsize}{!}{\includegraphics{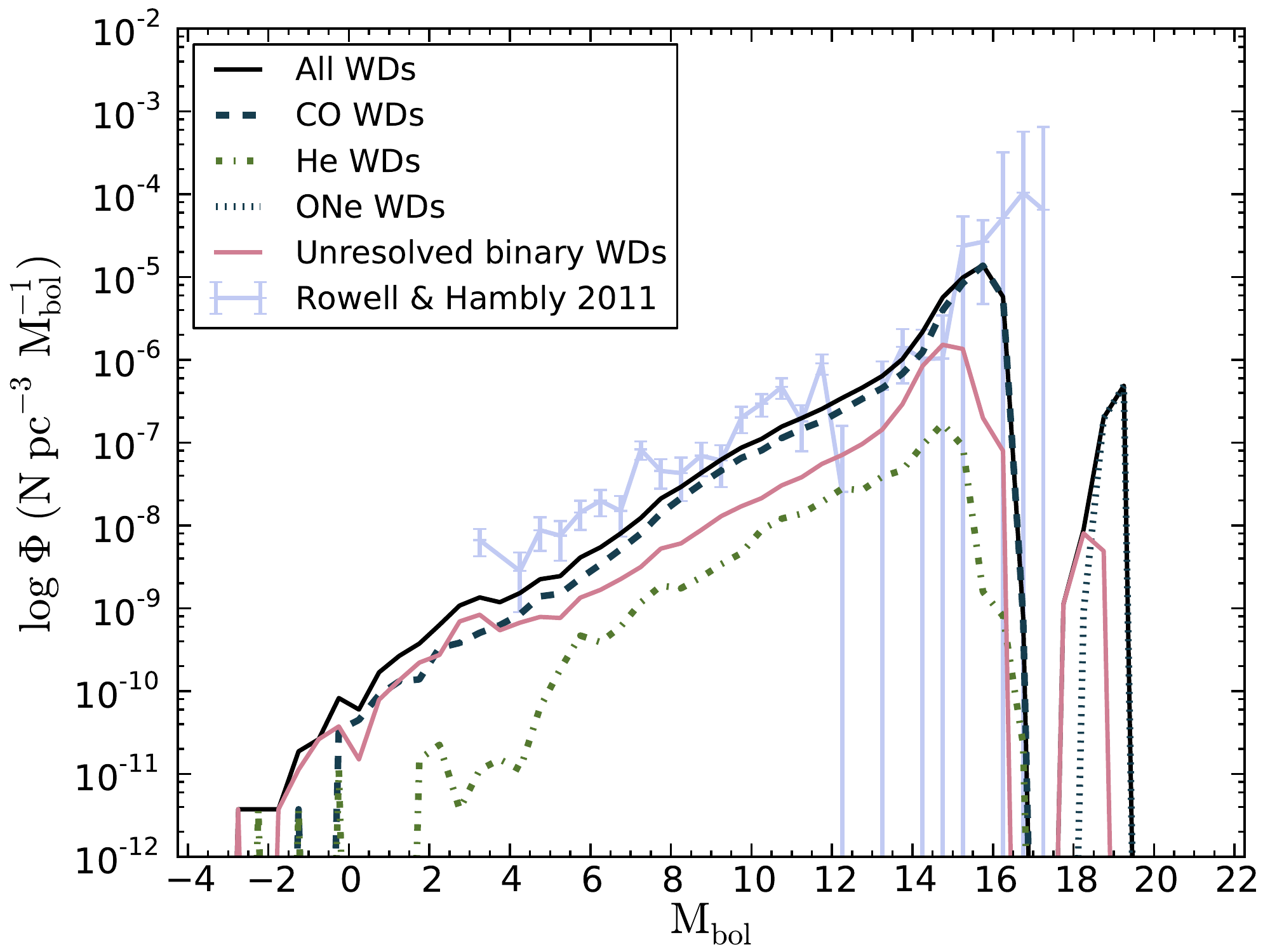}}
		\caption{Build-up of the WDLF in our standard halo model (50\% binaries). 
		All WDs are included in the black solid line (the total WDLF).
		The lower solid line shows the contribution
		from unresolved binary WDs. The dashed, dot-dashed and dotted lines
		show the contributions from CO, He and ONe WDs respectively.
		The light blue line with error bars is the halo WDLF
		determined from the SuperCOSMOS Sky Survey \citep{Rowell:2011},
		shown for comparison.}
		\label{07}
\end{figure*}

In section~\ref{sec:3} we have seen that the cooling tracks for He core WDs are quite different from those
with CO cores, which in turn differ from ONe core WD cooling tracks.
The effect of this can be seen partly in Figure~\ref{07}, 
where we present how the WDLF for our standard halo model
is built up from contributions of the various WD types.
We first note the monotonic increase in the WDLF, which occurs because 
the cooling of WDs is a simple gravothermal process \citep{Isern:2013}.
The drop in the number of stars at $M_\mathrm{bol} \approx 16$ is a consequence 
of the finite age of the universe. As was suggested by e.g. \citet{Winget:1987}, 
the observation of this drop can be used to constrain the age of our Galaxy. 
A second peak in the WDLF (the dotted curve around 
$M_\mathrm{bol} \approx 19$ in Figure~\ref{07}) is expected 
to consist of ONe WDs, due to their fast cooling times, 
as was first pointed out by \citet{Isern:1998}. 

The contribution from the He WDs to the WDLF is shown with a dot-dashed line
in Figure~\ref{07}. These He WDs have an unseen neutron star (NS) or black hole (BH) companion
or they are the resulting merger product of two stars in a binary.
However, there are many more He WDs that contribute to the WDLF:
those in unresolved binary WD pairs (the lower solid line in Figure~\ref{07}). 
Since they have slow cooling times,
the contribution of He WDs to the WDLF is largest at the bright end.
The unresolved binaries that end up in the second peak of the WDLF 
(around $M_\mathrm{bol} \approx 18.5$) are systems in which
at least one of the two WDs has an ONe core.
The main contributors to the WDLF are CO WDs, visualized with
a dashed line in Figure~\ref{07}, which is just below the black solid line.
These can be single CO WDs, CO WDs in wide binary WD pairs, but also
CO WDs with a NS or BH companion or merger products.
WDs with a main-sequence star as companion are not included
in the WDLF, because the light from the main-sequence star will
dominate the spectrum in that case.

Figure~\ref{07} shows that our standard model WDLF lies below
the observed WDLF (RH11; the light blue line with error bars 
in Figure~\ref{07}), however we shall see in the next subsection that
this discrepancy disappears when we vary the normalization.
Our integrated standard model luminosity function 
(the black solid line in Figure~\ref{07}) yields 
$n_\mathrm{Halo \ WDs} = 2.08 \cdot 10^{-5} \ \mathrm{pc}^{-3}$.
This value is lower than the integrated value of the RH11 WDLF, 
$(1.4 \pm 5.6) \cdot 10^{-4} \ \mathrm{pc}^{-3}$, 
mainly because of their higher estimate of the number of WDs 
in the luminosity bins around $M_\mathrm{bol} \approx 17$. 
Our models predict that there are no WDs in these bins.
Although the present-day estimate of the number of WDs with $M_\mathrm{bol} \approx 17$
should be regarded as an upper limit because of the large error bars,
future observations on the shape of this faint end of the WDLF will help
to constrain WD cooling models and the SFH of the halo, whereas the
normalization of the WDLF, especially at the faint end, will help 
to constrain the IMF and binary fraction (see section~\ref{comparing}).

When comparing our theoretically determined WDLF with the observed one by RH11
apart from the missing faint end (which is not reached by
SuperCOSMOS because it is a magnitude-limited survey), 
also the missing bright end catches the eye.
Here, another selection effect plays an important role:

RH11 only included WDs with a tangential velocity larger than $v_{t,\mathrm{min}} = 200$ $\mathrm{km} \ \mathrm{s}^{-1}$,
to filter out thin and thick disk WDs. Due to the mean lower proper motion completeness limit 
$\mu_\mathrm{min} = 40$ $\mathrm{mas} \ \mathrm{yr}^{-1}$ across the SuperCOSMOS Sky Survey,
the sample of RH11 is becoming less complete at a distance of approximately
\begin{equation}
\frac{d_\mathrm{max}}{\mathrm{pc}} = \left(\frac{p_\mathrm{min}}{\mathrm{arcsec}}\right)^{-1} = 
\frac{v_{t,\mathrm{min} \ }}{4.74 \ \mathrm{km} \ \mathrm{s}^{-1}} \left(\frac{\mu_\mathrm{min}}{\mathrm{arcsec} \ 
\mathrm{yr}^{-1}} \right)^{-1} \approx 1 \cdot 10^3. \
\end{equation}
Here, $p_\mathrm{min}$ is the minimum parallax that is reached
and we have used that a proper motion of $1 \ \mathrm{arcsec} \ \mathrm{yr}^{-1}$ 
corresponds to a tangential velocity of 
$1 \ \mathrm{AU} \ \mathrm{yr}^{-1} = 4.74 \ \mathrm{km} \ \mathrm{s}^{-1}$ at 1 pc.
Because at distances larger than $\sim 1$ kpc, young and bright halo WDs contribute more
to the WDLF than fainter WDs, the bright end of the WDLF is not reached by SuperCOSMOS.

\begin{figure*}
\centering	
	\includegraphics[width=0.47\textwidth]{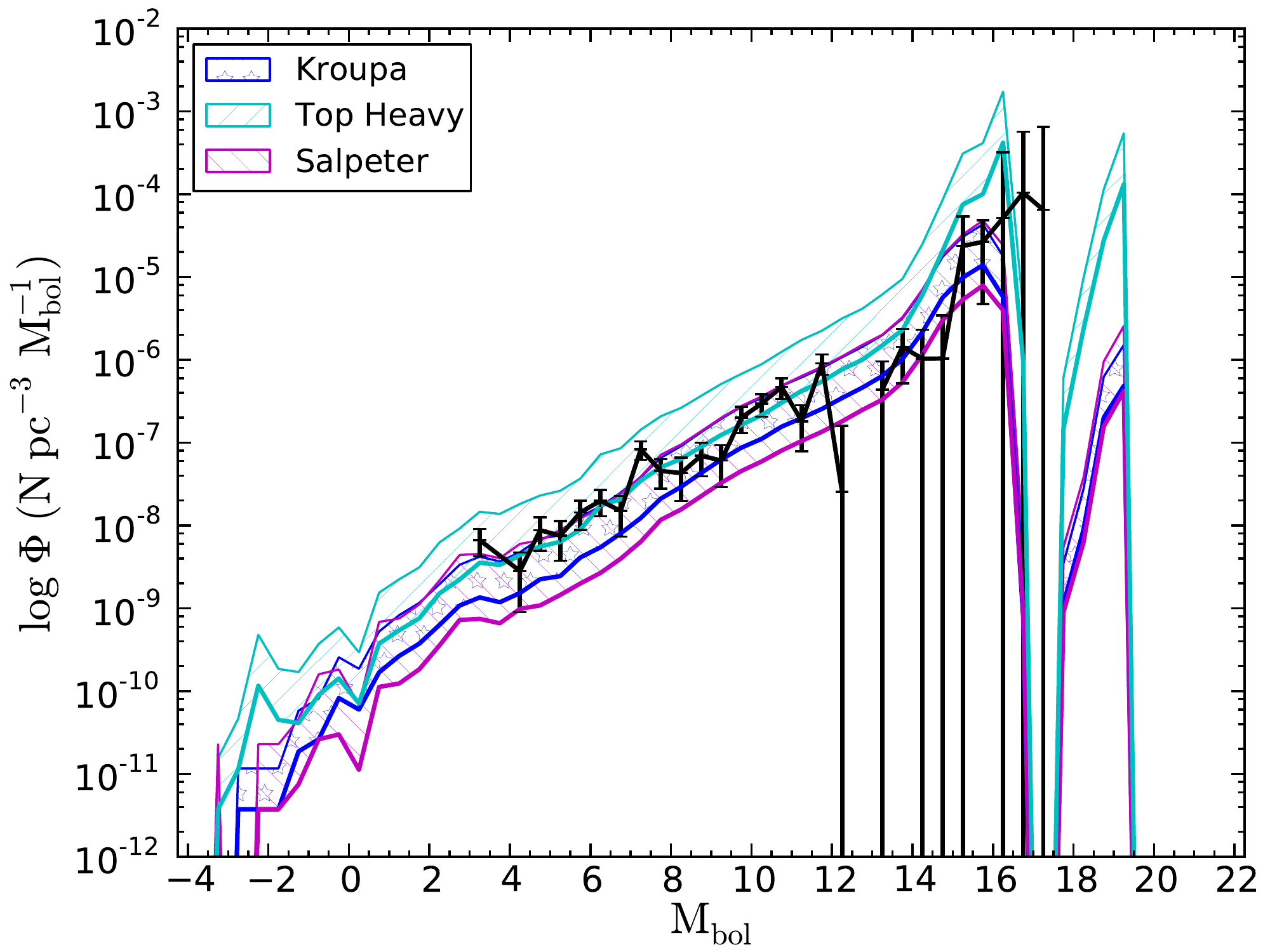}
	\includegraphics[width=0.47\textwidth]{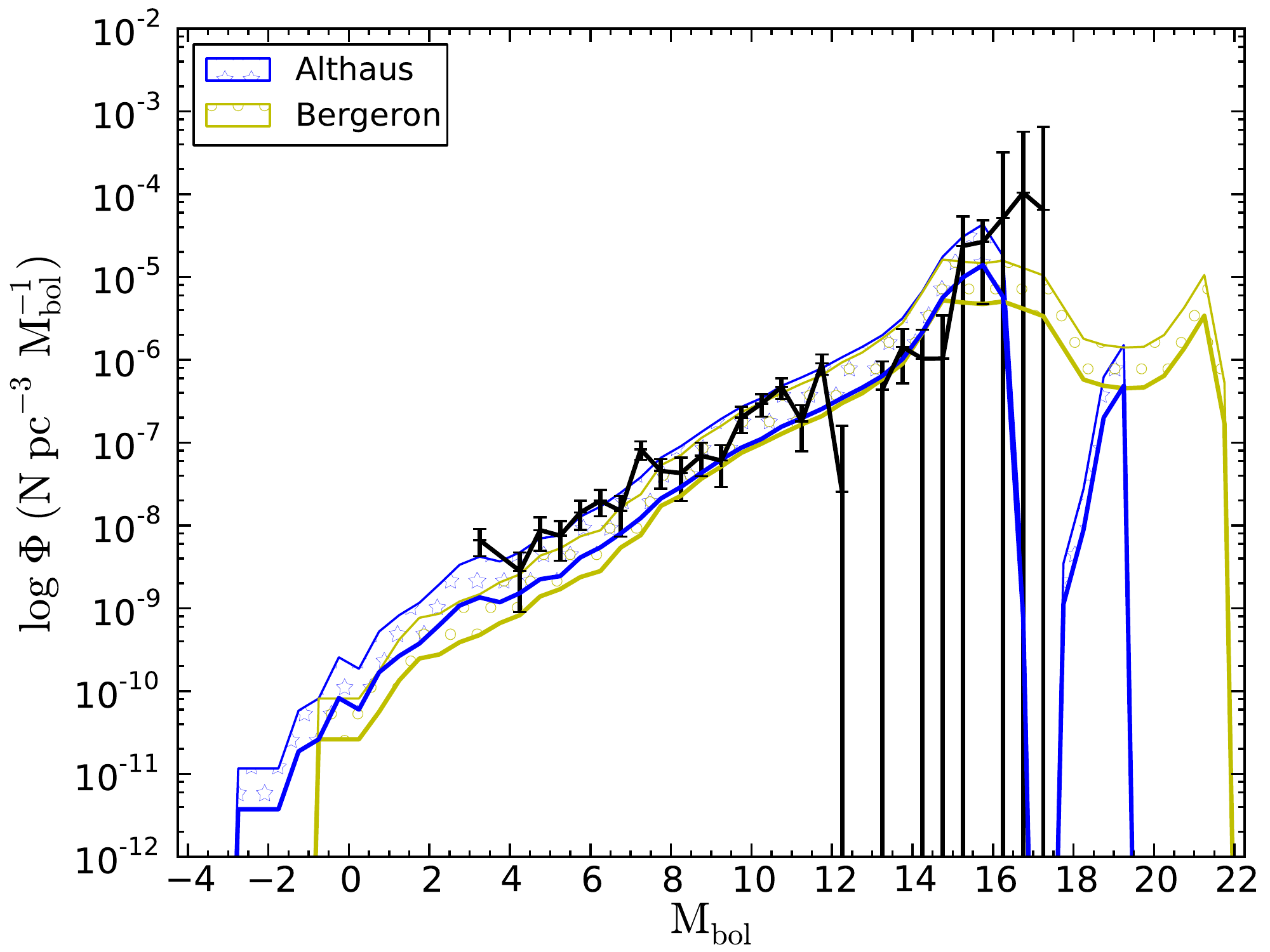}
	\includegraphics[width=0.47\textwidth]{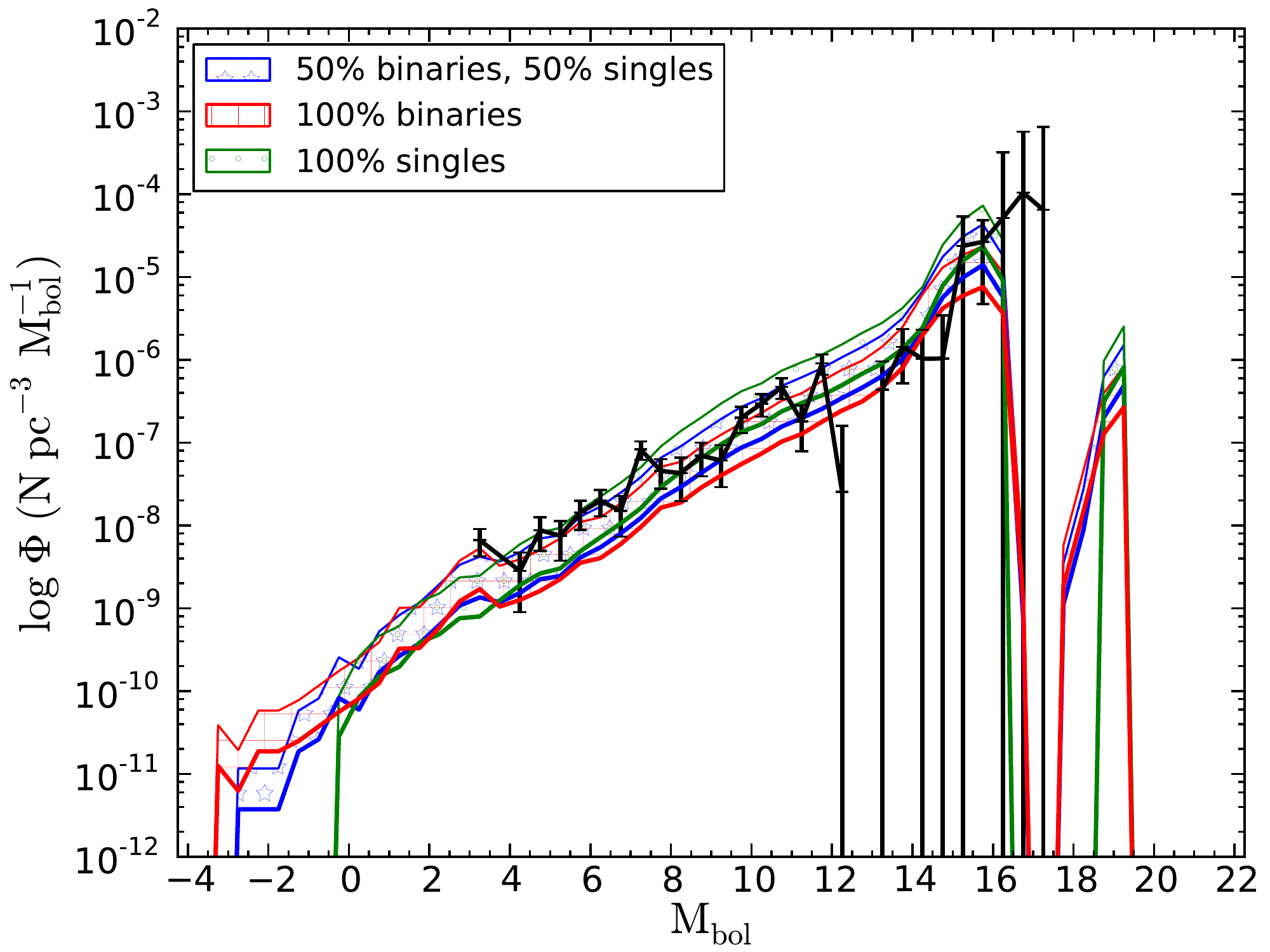}
	\includegraphics[width=0.47\textwidth]{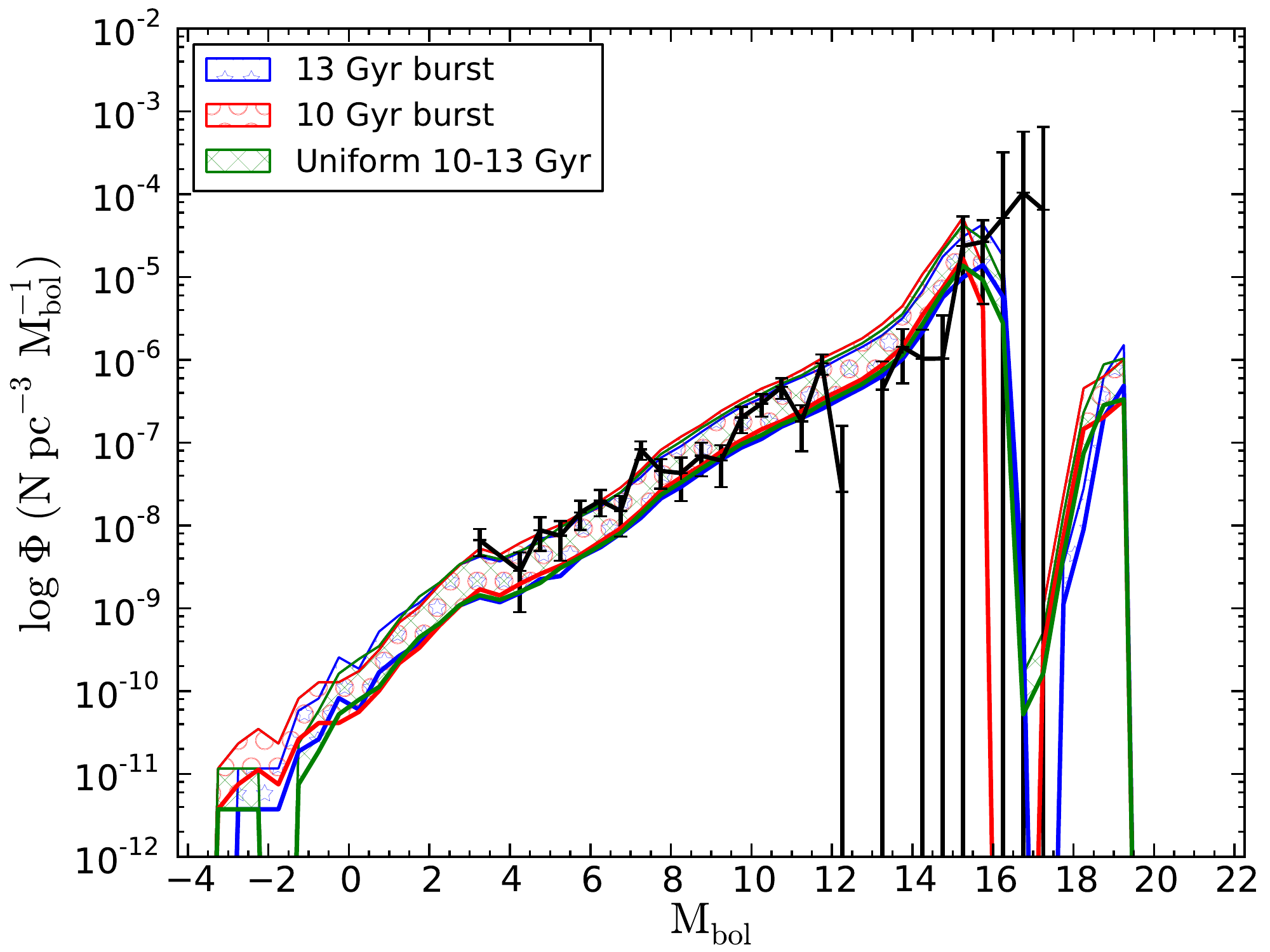}
		\caption{Comparison of halo WD luminosity functions corresponding to different assumptions about the IMF
		(top left), WD cooling (top right), binary fraction (bottom left) and SFH of the halo (bottom right).
		The WDLF corresponding to our standard model, indicated by the blue solid line in all panels, is 
		constructed using the $1.2 \cdot 10^7$ WDs in our simulation box (see Table~\ref{01}).}
		\label{06}
\end{figure*}

We expect this latter bias to be resolved by Gaia, which
can do microarcsecond astrometry and, as we will show in section~\ref{sec:7}, 
is expected to detect intrinsically bright WDs to distances of $\sim 2.5 \ \mathrm{kpc}$.
Although the bright end of the WDLF has already be determined from an
empirical measure of the WD cooling rate in a globular cluster (Goldsbury et al. 2012),
we will soon have a new window on the Galactic halo, 
when we start to explore bright field halo WDs with Gaia.

\subsection{Comparing the WDLFs of different halo models} \label{comparing}

Eight different model WDLFs are visualized in Figure~\ref{06}, comparing the effect of a different IMF,
cooling track, binary fraction and SFH.
As explained at the beginning of section~\ref{models}, we calculate not only the shape 
of the WDLF, but also derive its normalization from an independent mass density estimate of halo stars.
For each model in Figure~\ref{06} a band is given rather than a single line, which comes from the two different normalizations
explained in section~\ref{IMFs}. To arrive at the upper lines, the lower line is simply multiplied with
a normalization factor corresponding to the used IMF ($5.9/1.9$ for the Kroupa IMF, $6.7/1.1$ for the Salpeter IMF
and $330/80$ for the top-heavy IMF, see Table~\ref{table:01}).
The blue band in each panel represents our standard model, labelled ``Kroupa'', ``Althaus'', 
``50\% binaries, 50\% singles'' and ``13 Gyr burst'' respectively.

The left panels of Figure~\ref{06} show that a different IMF or binary fraction 
affects the normalization of the WDLF, as expected from sections \ref{IMFs} and \ref{2.2}.
Regarding the shape of these five different WDLFs in the left panels of Figure~\ref{06},
the differences are the largest at the extremely faint and the extremely bright end. 
We see that the WDLF corresponding to a model with a larger binary fraction 
resembles more closely the shape of the lower solid line
in Figure~\ref{07}, where the contribution to the WDLF from unresolved binary WDs is shown.

From the top right panel in Figure~\ref{06} it is clear that there is a significant difference
in the shape of the faint end of the WDLF when a different assumption is made about WD cooling.
The drop between the two peaks of the WDLF is less prominent when the Bergeron models
are used compared to the Althaus models, because CO WDs with a
luminosity $\log (L/L_\odot) < -4$ cool faster in the Bergeron models than in the Althaus models
(see the right panel of Figure~\ref{02}).

The logaritmic scale on the vertical axis implies that it will be observationally
challenging to distinguish the three different models of SFH (shown in the bottom right panel).
These differ slightly from each other at the the faint end of the WDLF.
As expected, the WDLF of a 10 Gyr halo drops off at lower magnitudes than a 13 Gyr old halo. 
Furthermore, the gap between the two peaks of the WDLF is more prominent 
in the models with a SF burst compared to models with a continuous SFH.

There is a quite good overall agreement between our theoretically
predicted WDLFs and the observed one by RH11, except for the model with a
top-heavy IMF, which overpredicts the number of WDs per luminosity bin at the faint 
end of the WDLF. This also follows from the reduced $\chi^2$-test that we conducted
to compare the agreement between the different model WDLFs in Figure~\ref{06} with 
the observationally determined WDLF quantitatively (see Table~\ref{table:04}).

\begin{table}
\caption{Reduced $\chi^2$ values for eight halo models.}
\begin{tabular}{ccccc}
\toprule 
Model &  $\chi^2$ & $\chi^2_\mathrm{upper}$ & $\chi^2_\mathrm{min}$ & $f$\\
\midrule
 Standard &	   $2.29$ & $7.07$ & $2.26$ & $1.28$	\\
\midrule
 Salpeter &    $2.99$ & $7.76$ & $2.27$ & $2.40$	\\
\midrule
 Top-heavy & 	$5.74$ & $140$ & $2.67$ & $0.44$	\\
\midrule
 Bergeron &     $2.74$ & $5.68$ & $2.68$ & $1.38$ \\
 \midrule
 100\% singles &    $2.28$ & $16.8$ & $2.33$ & $0.88$ \\
\midrule
 100\% binaries & 	$2.61$ & $3.57$ & $2.25$ & $1.76$ \\
 \midrule
 10 Gyr burst & 	$2.35$ & $13.5$ & $2.43$ & $0.95$	\\
\midrule
 Uniform 10$-$13 Gyr & 	$2.31$ & $9.50$ & $2.37$ & $1.12$\\
\bottomrule
\end{tabular}
\tablefoot{\small Reduced $\chi^2$ values for all model WDLFs (first column),
reduced $\chi^2$ values corresponding to the upper limits (second column),
minimum value that the reduced $\chi^2$ can become ($\chi^2_\mathrm{min}$)
by multiplying the model WDLF with a factor $f$ (third and fourth column).
In the first two columns there are 28 degrees of freedom, in the last 
column there is one degree of freedom less.}
\label{table:04}
\end{table}

From the first column of Table~\ref{table:04} we see that our standard model and the
model with 100\% singles have the lowest reduced $\chi^2$ values ($\chi^2 = 2.29$ and $2.28$
respectively), closely followed by the models with alternative SFHs 
($\chi^2 = 2.31$ for the model with uniform SF
between 10 and 13 Gyr, and $\chi^2 = 2.35$ for the model with a SF burst 10 Gyr ago). 
The fact that all these values are so close together can also be determined from Figure~\ref{06}, 
where these four curves almost completely overlap.
The models with 100\% binaries or a Salpeter IMF do slightly worse, due to their lower
normalization.
In all cases the line corresponding to the upper limit of the number of stars 
has worse agreement with the observed WDLF ($\chi_\mathrm{upper}^2$; second column) 
than the lower line corresponding to that same model.
This seems to indicate that the low-mass part of the IMF does not turn over at $\sim 1.0 \ M_\odot$
to become completely flat, but rather has a negative slope.

We varied the normalization of the WDLFs by multiplying them with a free parameter $f$,
to see how well we can fit the shape of the WDLF. We kept $f$ as a free parameter,
because there are many ways in which we could adapt the normalization, for example choosing a
different $\gamma_\mathrm{unev}$ (as we did for calculating the upper limits), a different
binary fraction or a different mass density in unevolved stars $\rho_0$. 
The results of this analysis (summarized in the parameter $\chi_\mathrm{min}^2$) 
are given in the third column of Table~\ref{table:04} 
for each model with the corresponding $f$ value in the fourth column. 
Without normalizing the model WDLFs, 
the model with 100\% binaries comes out best, with a reduced $\chi_\mathrm{min}^2$ value of 2.25.
Although these minimum $\chi_\mathrm{min}^2$ values lie close together for most of the models,
the Top-heavy and Bergeron models still have the worst agreement with the WDLF observed by RH11.
For the two models with alternative SFHs and the model with 100\% singles
the $\chi_\mathrm{min}^2$ values are larger than the $\chi^2$ value corresponding to 
our preferred normalization, because there is one degree of freedom less if we fix the normalization of the WDLF.

The $\chi^2$ values are also affected by our assumption of $P(v_\mathrm{t} > 200)$.
If we had chosen the value $P(v_\mathrm{t} > 200) = 0.4$ or 0.5, our standard model
$\chi^2$ value would change to 2.39 or 2.22 respectively, and how this other choice
affects the other curves can be determined from the parameter $f$. If $f$ is larger than 1,
the larger value $P(v_\mathrm{t} > 200) = 0.5$ would reduce the $\chi^2$ value,
if $f$ is smaller than one $P(v_\mathrm{t} > 200) = 0.4$ would yield a better match.

\subsection{Halo white dwarfs detectable by Gaia} \label{sec:7}

In this subsection of our results, we take a closer look at the
population of halo WDs in our standard model and what fraction of this
population can be seen by the Gaia satellite. 

\begin{figure} 
\centering	
	\resizebox{0.8\hsize}{!}{\includegraphics{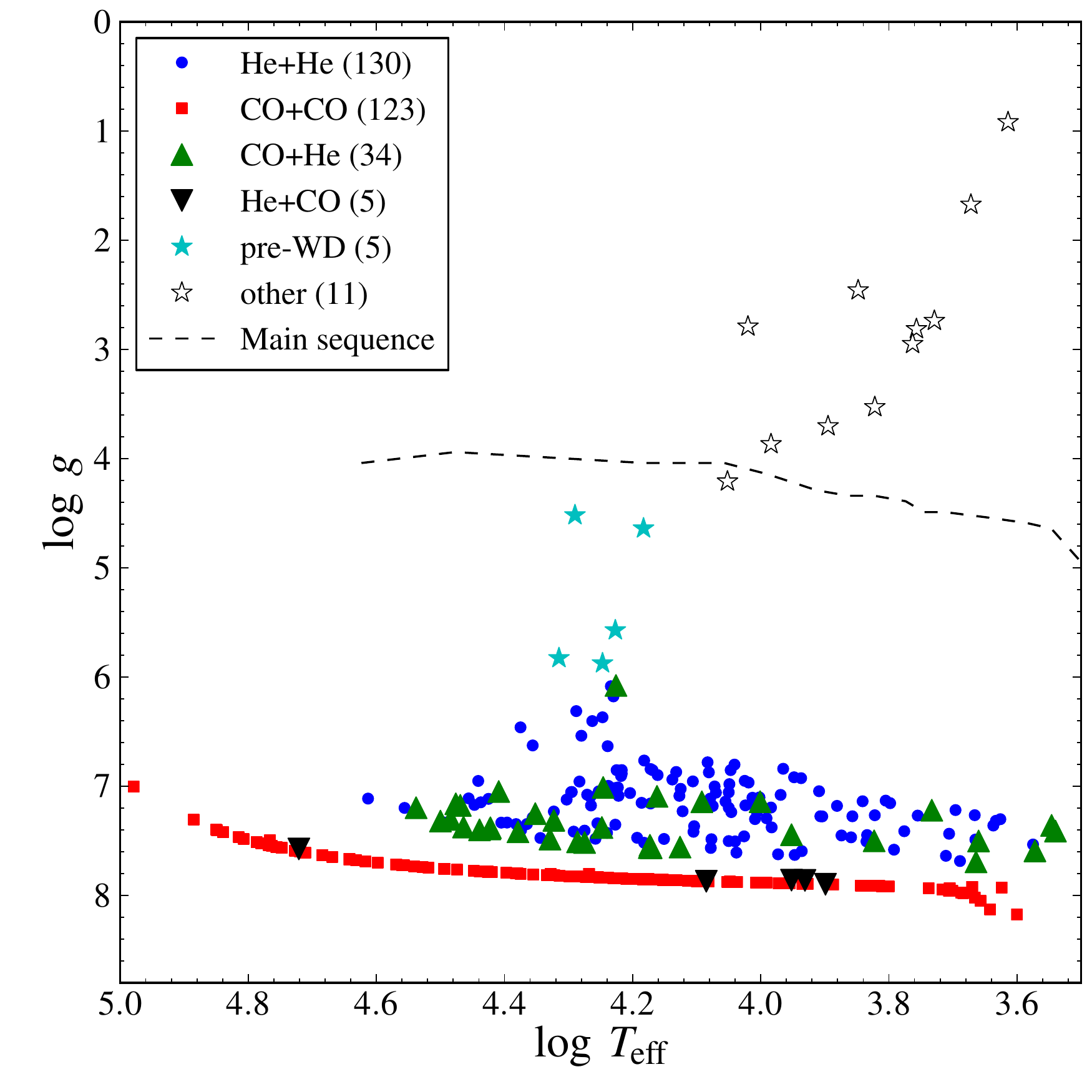}}
	\resizebox{0.8\hsize}{!}{\includegraphics{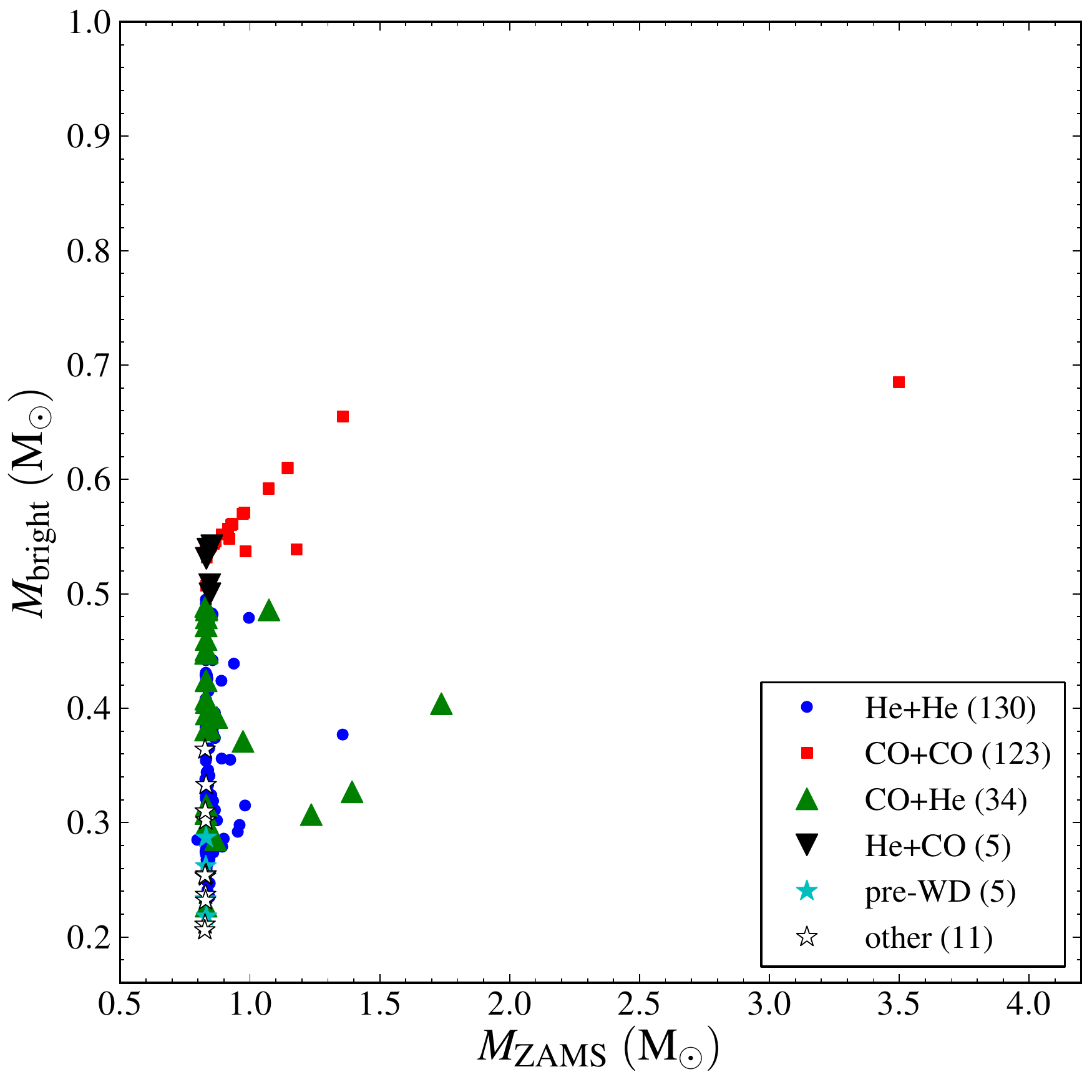}}	
	\resizebox{0.8\hsize}{!}{\includegraphics{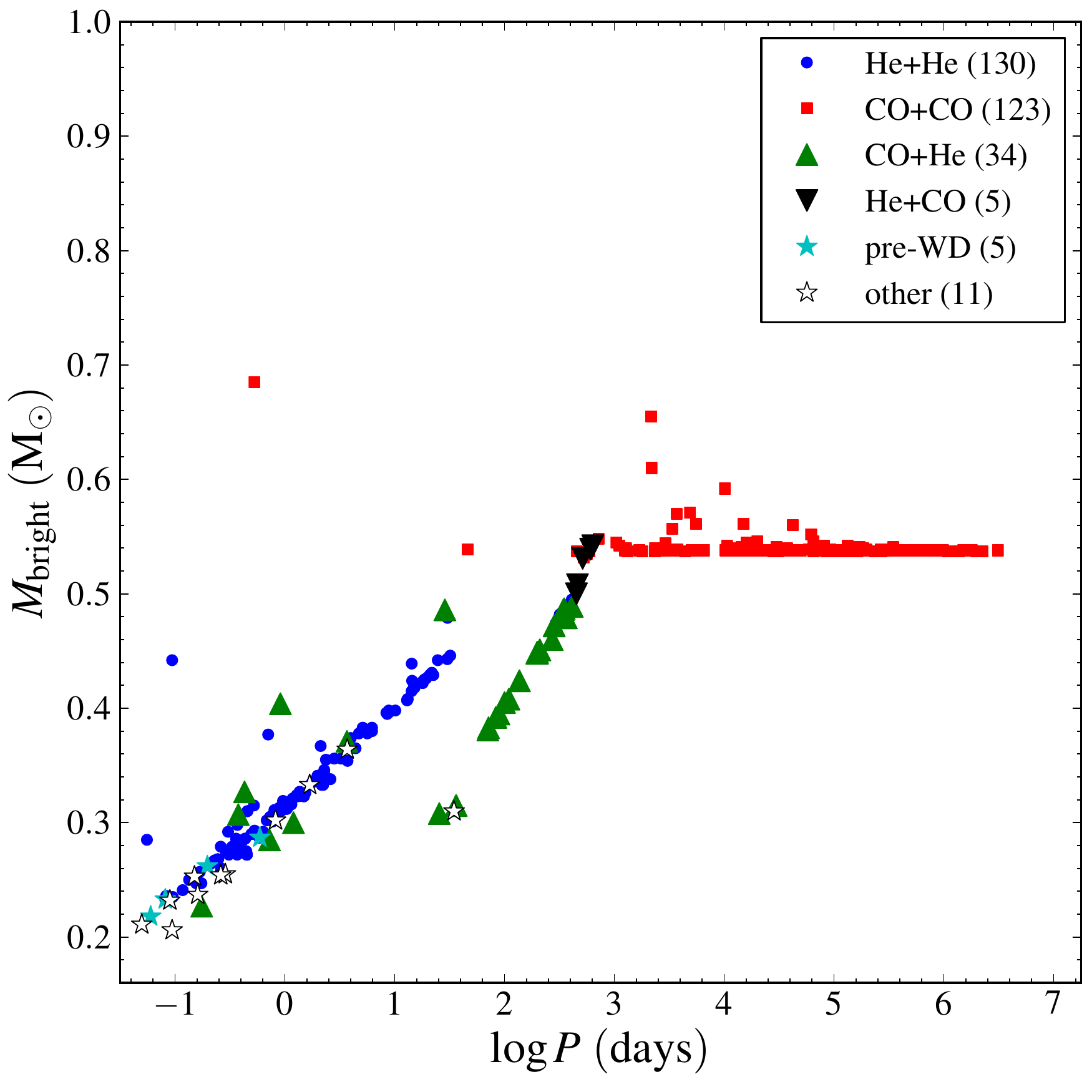}}
	\caption{Properties of unresolved binary WD pairs with $G<20$
	for our standard model. Shown are double He WDs (circles), double CO WDs 
	(squares), CO+He systems in which the He WD is the brightest
	(upward pointing triangles), He+CO systems in which the 
	CO WD is the brightest (downward pointing traingles), pre-WDs (filled stars),
	and ``other'' stars, which are pre-WDs that are indistinguishable from MS stars or giants.
	After the label discriptions in the legend, the total number of WD binaries 
	of that particular kind is given. 
	Due to the low number of halo WDs with $G<20$ we expect to find in the
	Milky Way, there is some statistical noise in this Figure.}
	\label{04}
\end{figure}

\begin{figure} 
\centering	
	\resizebox{0.8\hsize}{!}{\includegraphics{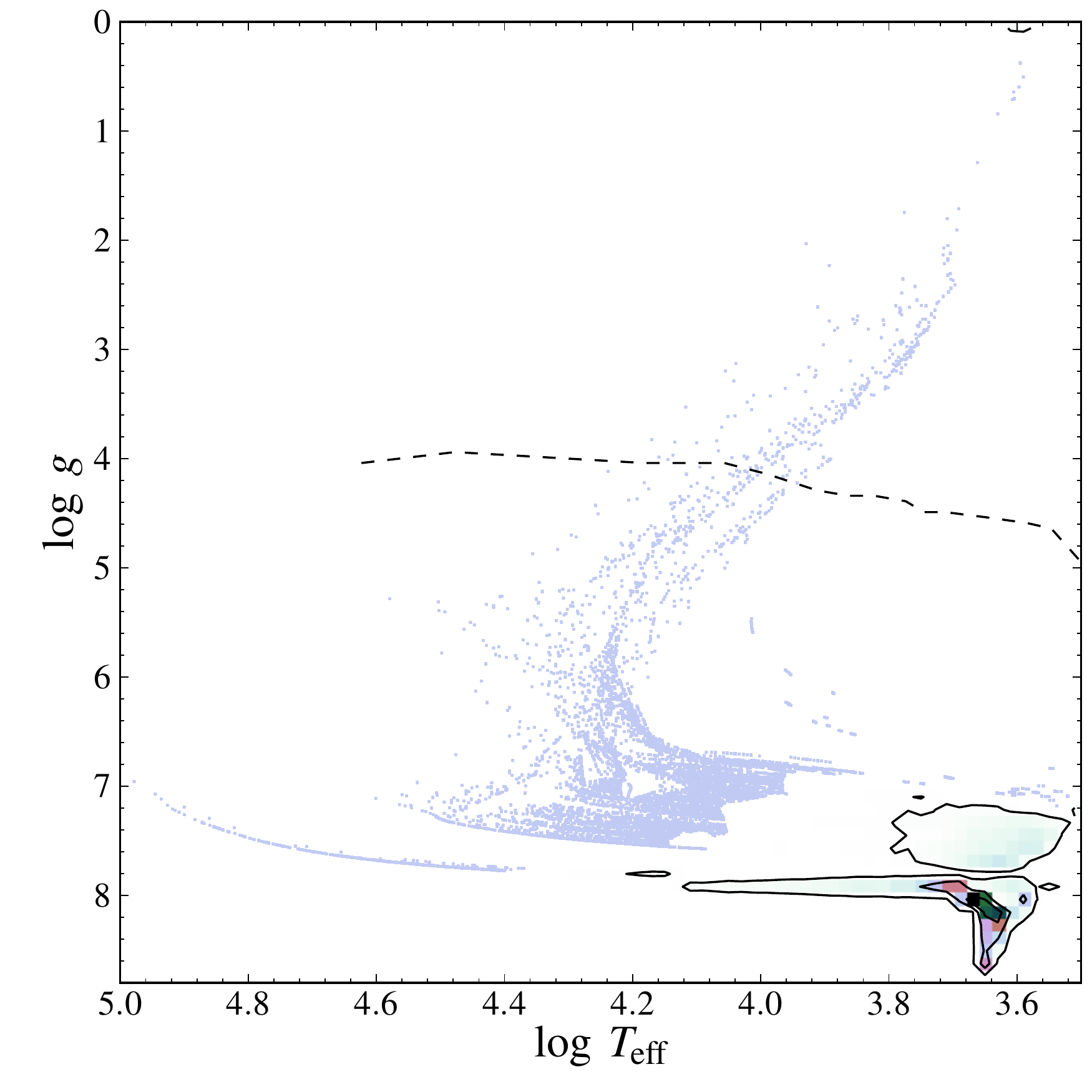}}
	\resizebox{0.8\hsize}{!}{\includegraphics{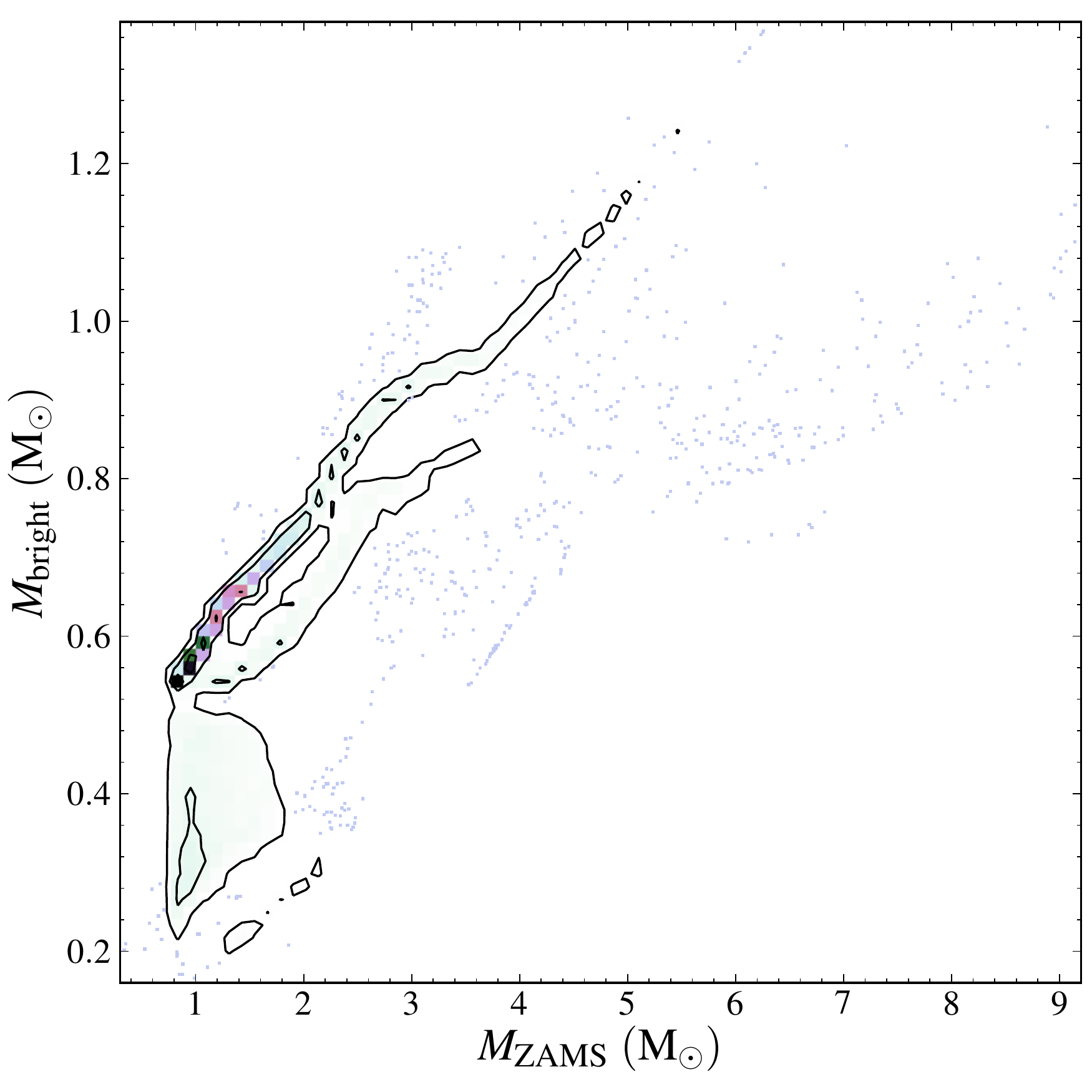}}
	\resizebox{0.8\hsize}{!}{\includegraphics{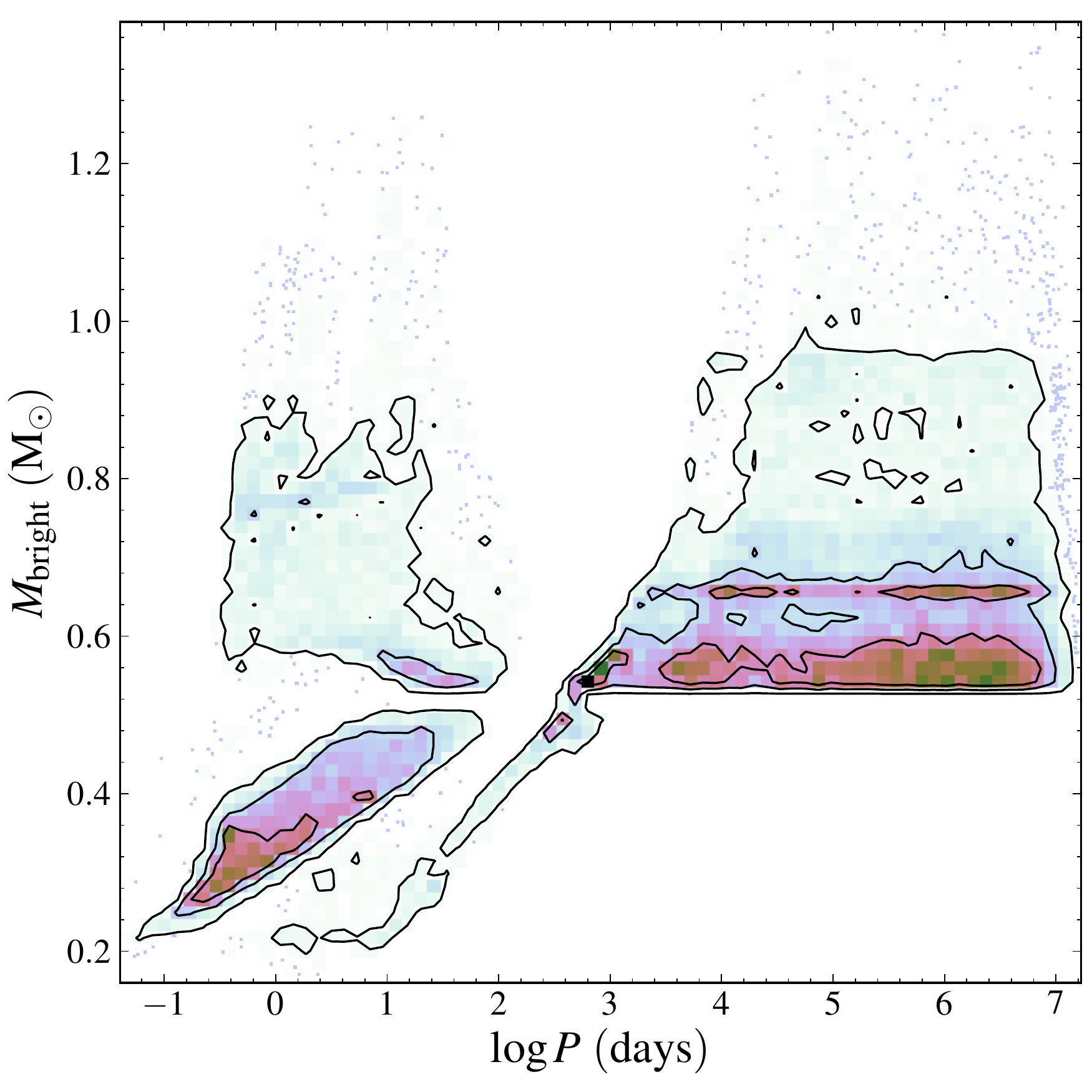}}
	\caption{Properties of all $1.3 \cdot 10^6$ unresolved binary WD pairs 
	in our simulation box (maximum magnitude $G=35$), for our standard model.
	Contour lines mark the regions in which 33\%, 67\% and 95\% of the binaries are located.
	Higher density regions are given darker colours. The 1\% binaries in the tail of the distribution 
	are indicated by the scattered points. Compared to Figure~\ref{04}, this Figure 
	hardly has statistical noise because of the large number of halo WDs we expect
	to find in the Milky Way halo if we would go up to magnitude $G=35$.}
	\label{05}
\end{figure}

An important point to keep in mind when studying halo WDs,
is that one is biased towards young and bright WDs in a magnitude-limited survey.
Since the bright part of the WDLF is to a large extent built up by unresolved binary WD pairs 
(see Figure~\ref{07}), we first look at their properties.
Figure~\ref{05} shows the properties of all unresolved binary WD pairs in our simulation 
box, whereas Figure~\ref{04} focusses on the $\sim 300$ unresolved binary WD pairs
with $G<20$. We note that this also includes binaries with large orbital separations 
which have never undergone interaction, because at large distances these can still be unresolved.
Of the two WDs in each binary, the properties of the brightest are plotted.
A distinction is made between CO+He WDs and He+CO WDs, the second of the 
two WD types in each group is the brightest WD in the system. For most of the systems,
this is also the youngest WD. However, in some CO+He systems the He WD was formed first
and is still brighter than the later formed CO WD, which is possible since
He WDs can be intrinsically brighter than CO WDs at birth and they in general
have longer cooling times (see section~\ref{sec:3}). 
In the legend of each panel in Figure~\ref{04},
the number of systems of that particlar kind is given in brackets.
Due to the low number of halo WDs we expect to find in the
Milky Way, there is some statistical noise in this Figure.
What we want to show here, are the global positions of the
WDs in this diagram, without focussing on their individual positions.

A particular aspect of the cooling models that we use as our standard, 
is that the luminosity of He WDs stays constant for a long time 
($10^5 - 10^9$ yr, depending on the mass), 
before cooling starts. This can be seen from the dark dashed line 
in the left panel of Figure~\ref{02}.
As a consequence of this feature, He WDs that are on this part the cooling track
will be seen more often than CO WDs with the same cooling time.
We will refer to them as pre-WDs, to indicate that these objects do
not look like standard WDs, because they are brighter and have smaller
surface gravities.

In the top panels of Figures~\ref{04} and \ref{05}, which are $\log g - \log T_\mathrm{eff}$ diagrams, 
the pre-WDs (plotted with star symbols) are clearly distinguishable from the other WDs because of their low surface gravities.
We define pre-WDs as those double WDs in which the brightest of the two has $4.3 < \log g < 6$.
There are more He WDs with even lower surface gravities,
indicated by the label ``other'' in Figure~\ref{04}. However, these will be hard to distinguish from main-sequence
stars or giants, which lie on or above the dashed line at $\log g \approx 4$ in Figure~\ref{04} \citep{Allen:1973}.
Because pre-WDs are only apparent when we use the Althaus cooling tracks
for He WDs (see Figure~\ref{02}), Figures~\ref{04}
and \ref{05} would not have any points with $\log g < 6$
if we use the models with Bergeron cooling instead of our standard model.
In the top panel of Figure~\ref{05} we see that more than 95\% of the halo
WDs have $\log g > 7.0$ and $\log T_\mathrm{eff} < 4.2$, which is also the part of the diagram where 
most of the single WDs and resolved WD binaries are expected to be situated.
Furthermore, we see from the top panel of Figure~\ref{04} that there is a narrow gap between
the $\log g$ and $\log T_\mathrm{eff}$ values of systems in which the brightest WD has a CO core
and those in which the brightest of the two has a He core.
In this way, these systems can in principle be distinguished from their positions
in the $\log g - \log T_\mathrm{eff}$ diagram.

The middle panels of Figures \ref{04} and \ref{05} show the IFMR for halo WDs, i.e.
the mass of the brightest WD in the binary system ($M_\mathrm{bright}$) 
is plotted as a function of its corresponding initial mass ($M_\mathrm{ZAMS}$). 
In Figure~\ref{04} we see that in most of these unresolved binary WDs 
the brightest of the two stars has a main-sequence progenitor star 
with a mass $M_\mathrm{ZAMS} \approx 0.84~\mathrm{M}_\odot$.
Because in our standard model halo stars are born about 13 Gyr ago,
these stars have just become WDs, thus will be very abundant in the Gaia catalogue
of halo WDs. There are very few high-mass WDs in our sample, mainly because their progenitor 
stars have shorter main-sequence lifetimes and they have thus cooled down much more.
In the middle panel of Figure~\ref{05}, the global IFMR for halo WDs is shown.
There is no focus on only the brightest WDs, with the result that
the highest density region is shifted towards a line that resembles the IFMR of single stars,
populated by the unresolved binary WDs that have not undergone interaction.

In the bottom panels of Figures~\ref{04} and \ref{05} the $M_\mathrm{bright} - $ orbital period relation 
for halo WDs is shown. We see that the unresolved binary WD pairs with $G<20$ lie on three 
distinct lines, where each line is mostly populated by one of the different binary types. 
The majority of double CO systems have not interacted and thus evolve to systems
with wide periods. Because most stars are low-mass they
form WDs with similar masses, while the periods are determined by the
initial period distribution. 
The short-period branch shows systems that are formed via
CE evolution. In our model, this CE between a giant and a WD
is always described by a the energy balance \citep[$\alpha$, see][]{Toonen:2012}.
The correlation they show between WD mass 
and orbital period can then be understood from the relation 
between the core mass and the radius of giants. 
Systems that start the CE phase in a more compact orbit will have giants with smaller radii 
and thus lower-mass cores. This means both a spiral in to shorter final periods
and a final WD mass that is lower. 
The branch with longer periods shows systems that are formed via
a second phase of mass transfer that was stable. During the mass
transfer the orbit widens, which stops when the whole envelope of the
giant has been transferred to the first formed WD. Due to the same
relation mentioned above, giants with larger core masses (that form more
massive WDs) are bigger and thus end their evolution in binaries with
longer orbital periods. The same relation is seen in the WD companions
to millisecond radio pulsars \citep{Savonije:1987}.
From Figure~\ref{05} it is clear that Figure~\ref{04} only resembles 
a small part of the complete parameter space, but it constitutes a representative
selection of the low-mass part of this diagram.

White dwarfs in unresolved binaries are of course not the only halo WDs we expect Gaia to observe.
As we already mentioned in section~\ref{sec:6}, single WDs, resolved double WDs,
WDs with a NS or BH companion, and WDs that are the result of a merger also contribute to the WDLF.
The number of WDs in each of these five groups is specified per
WD type (pre-WD, He, CO, or ONe core) in Table~\ref{table:03}.
We see that all single WDs and all brightest WDs in a resolved binary
system have a CO core with the limiting magnitude 
$G<20$. If we look at fainter magnitudes, e.g. $G<23$ 
or $G<26$, Table~\ref{table:03} shows that ONe WDs 
will be detected, although there are still very few of them compared to CO WDs. 
The same is true for WDs with a NS or BH companion.

When selecting halo WDs from the Gaia catalogue, selection effects are expected, like
the factor $P(v_\mathrm{t} > 200)$ that we multiplied our theoretically 
determined WDLFs with in the previous subsections to compare our results with that of RH11.
Here, we do not include this factor, since it is not yet clear how large it will be.
For example, for some fraction of the stars ($V< 17$), radial velocities will also be available. 
Therefore it should be possible to obtain a larger number of halo WD stars than just 
with a cut in $v_\mathrm{t}$. Furthermore, the determination of the initial number of stars in our
simulation box has a greater effect on the number of halo WDs than these selection effects have.

In the top and bottom rows of Table~\ref{table:03}, the total number of halo WDs 
in two spheres around the Sun with respective radii 400 pc and 2.95 kpc is given,
as well as the total mass these halo WDs constitute.
From these we calculate the number densities of halo WDs $n_\mathrm{Halo \ WDs} = 5.89 \cdot 10^{-5} \ \mathrm{pc}^{-3}$ 
(within 400 pc) and the slightly higher value $n_\mathrm{Halo \ WDs} = 6.00 \cdot 10^{-5} \ \mathrm{pc}^{-3}$ 
(within 2.95 kpc). These values are more than a factor of two lager than the number density we derived
by integrating the WDLF in section~\ref{sec:6}.
This difference is due to the factor $P(v_\mathrm{t} > 200)$ which is not taken
into account here. Furthermore, here all halo WDs are counted within spheres of a certain radius around the Sun,
whereas in section~\ref{sec:6} we estimated the number density including the edges of our simulation box
(which is not a sphere, see Figure~\ref{01}).

\begin{table*}
\caption{Number of halo WDs in our simulation box.}
\begin{tabular}{cccccccc}
\toprule 
				& 								& single WD			& merger$\to$WD		& WD$+$WD unresolved& WD$+$WD resolved 	&  WD$+$NS / BH & Total\\
\midrule
				& $N_{\mathrm{total},G<20}$ 		& 395 				& 83				& 85 				& 58				& --- 			& 621\\ 
$d<400$ pc 		& $N_\mathrm{total}$			& $1.04 \cdot 10^4$	& 2604				& 1340				& 1447				& 29			& $1.58\cdot 10^4$\\
				& $M_\mathrm{total}$			& $7.10 \cdot 10^3$	& $1.83 \cdot 10^3$	& $1.63 \cdot 10^3$	& $2.09 \cdot 10^3$	& 30.0			& $1.27\cdot 10^4$\\
\midrule
				&$N_\mathrm{pre-WD}$			& ---   			& 1					& 5					& ---				& --- 			& 6\\
				&$N_\mathrm{brightest = He}$	& ---   			& 99  				& 164				& ---				& --- 			& 263\\
$G<20$	&$N_\mathrm{brightest = CO}$	& 872				& 97				& 128				& 157				& --- 			& 1254\\ 
				&$N_\mathrm{brightest = ONe}$ 	& --- 				& ---   			& ---				& --- 				& --- 			& ---\\
\cmidrule(l){2-8}
				&$N_\mathrm{total}$				& 872 				& 197				& 297 				& 157				& --- 			& 1523\\ 
\midrule
				&$N_\mathrm{pre-WD}$			& ---				& 29 				& 365				& --- 				& ---			& 394 \\
				&$N_\mathrm{brightest = He}$	& ---				& $4.63\cdot 10^3$	& $7.00\cdot 10^3$	& --- 				& ---			& $1.16\cdot 10^4$\\ 
$G<23$	&$N_\mathrm{brightest = CO}$	& $3.72\cdot 10^4$	& $4.57\cdot 10^3$	& $6.35\cdot 10^3$	& $5.17 \cdot 10^3$	& 2				& $5.33\cdot 10^4$\\ 
				&$N_\mathrm{brightest = ONe}$	& 5					& --- 				& ---				& ---				& 1				& 6\\
\cmidrule(l){2-8}
				&$N_\mathrm{total}$				& $3.73\cdot 10^4$	& $9.23\cdot 10^3$	& $1.37\cdot 10^4$	& $5.17 \cdot 10^3$	& 3 			& $6.54 \cdot 10^4$\\ 
\midrule
				&$N_\mathrm{pre-WD}$			& ---				& 33 				& 377				& --- 				& ---			& 410\\
				&$N_\mathrm{brightest = He}$	& ---				& $4.56\cdot 10^4$	& $7.65\cdot 10^4$	& --- 				& 53			& $1.22\cdot 10^5$\\ 
$G<26$	&$N_\mathrm{brightest = CO}$	& $5.99\cdot 10^5$	& $1.00\cdot 10^5$	& $1.28\cdot 10^5$	& $8.97\cdot 10^4$	& 145			& $9.18\cdot 10^5$\\ 
				&$N_\mathrm{brightest = ONe}$	& 180				& 28 				& 12				& 13				& 9				& 242\\
\cmidrule(l){2-8}
				&$N_\mathrm{total}$				& $6.00\cdot 10^5$	& $1.46\cdot 10^5$	& $2.06\cdot 10^5$	& $8.97\cdot 10^4$ 	& 207 			& $1.04 \cdot 10^6$\\  
\midrule
			   &$N_\mathrm{pre-WD}$			& ---				& 21				& 405				& --- 				& ---				& 426	\\
$d<2.95$ kpc   &$N_\mathrm{brightest = He}$	& ---				& $9.21\cdot 10^4$  & $1.66\cdot 10^5$	& --- 				& 528				& $2.58 \cdot 10^5$\\
		       &$N_\mathrm{brightest = CO}$	& $4.15\cdot 10^6$	& $9.60\cdot 10^5$	& $5.98\cdot 10^5$	& $3.78 \cdot 10^5$	& $6.87\cdot 10^3$ 	& $6.10 \cdot 10^6$	\\
			   &$N_\mathrm{brightest = ONe}$	& $7.39\cdot 10^4$	& $1.61\cdot 10^4$	& $2.23\cdot 10^3$	& $2.05 \cdot 10^3$	& $2.62\cdot 10^3$	& $9.69 \cdot 10^4$	\\ 
\cmidrule(l){2-8}
				&$N_\mathrm{total}$				& $4.23\cdot 10^6$ 	& $1.07\cdot 10^6$ 	& $7.66 \cdot 10^5$	& $3.80 \cdot 10^5$ & $1.00\cdot 10^4$ 	& $6.45 \cdot 10^6$\\ 
				&$M_\mathrm{total}$				& $2.89\cdot 10^6$	& $7.49\cdot 10^5$	& $9.76 \cdot 10^5$ & $5.54 \cdot 10^5$ & $9.47\cdot 10^3$  & $5.18 \cdot 10^6$\\
\bottomrule				
\end{tabular}
\tablefoot{The three middle rows indicate magnitude-limited selections 
of halo WDs in our simulation box.
For the volume-limited selections ($d<400$ pc and $d<2.95$ kpc) 
the total mass in WDs is indicated by $M_\mathrm{total}$.
These numbers are determined using our standard model (50\% binaries).
A long dash (---) indicates that the particular combination does not occur.}
\label{table:03}
\end{table*}

\begin{figure*}
\centering	
	\resizebox{0.78\hsize}{!}{\includegraphics{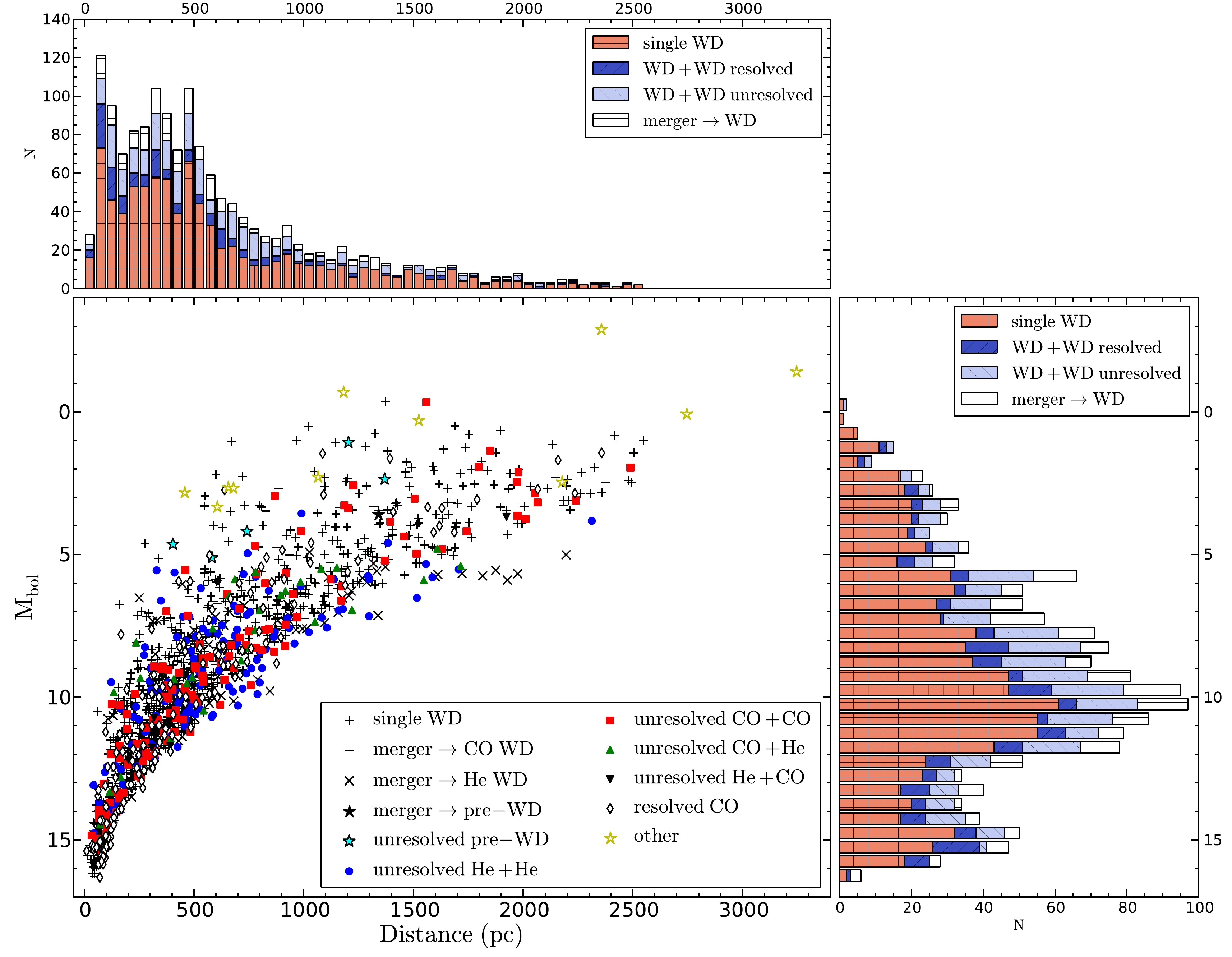}} 
		\caption{Distances and bolometric magnitudes of the $1.5 \cdot 10^3$ halo WDs 
		that can be observed with Gaia, for our standard model. Top panel: distribution of their distances.
		Right panel: distribution of their bolometric magnitudes, which gives an idea of the		
		statistical errors that are to be expected per luminosity bin of the WDLF for Gaia.
		The yellow stars, labelled ``other'', are pre-WDs that are indistinguishable 
		from MS stars or giants. These are not included in the
		projected distribution histograms.}
		\label{08}
\end{figure*}

\citet{Torres:2005} also estimated the number of (single) halo WDs with $G<20$ within 400 pc,
and found 542 (their Table~3). We found slightly more halo WDs within 400 pc:
621, including both singles and binaries, see the top row of Table~\ref{table:03}.
However, it is not strange that these numbers differ from each other, given the large number
of uncertainties in our estimate of the number density of halo WDs from the
observed mass density in unevolved stars (section~\ref{IMFs}) and the selection effects.
The latter are implicitly taken into account by \citet{Torres:1998}, because they normalized 
the number density of halo WDs within 400 pc to the local observed value \citep{Torres:1998}.

To check that our simulation box is large enough and make the claim that Gaia can detect
approximately $1.5 \cdot 10^3$ halo WDs with $G<20$ (Table~\ref{table:03}),
we plotted the distances to all these WDs as a function of their bolometric magnitude 
in Figure~\ref{08}. With the same markers as in Figure~\ref{04} the unresolved binaries
are visualized, additional markers are used for single WDs, resolved binary WDs
and WDs that are merger products. We see that apart from a few outlying ``other'' WDs,
all WDs fit well within the sphere with radius $\xi=2.95$~kpc around the Sun, which
validates the size of our simulation box. As explained above, the ``other'' WDs were
excluded from our luminosity function anyway because of their low surface gravities.

Projected onto the vertical axis of Figure~\ref{08} is the number of halo WDs
that can be found in every luminosity bin of the WDLF from the Gaia
catalogue. From this right panel of Figure~\ref{08} we get an idea of the
statistical errors that are to be expected per luminosity bin of the WDLF.
We see that the faint end of the WDLF will stay underdetermined 
since we expect to detect only a handful of WDs with $M_\mathrm{bol} > 16$ with Gaia
in our standard halo model. 
However, already with the few WDs in the lowest luminosity bins that can be reached,
we can start comparing our halo models. It is not clear whether
the drop at $M_\mathrm{bol} \approx 16$ is a detection limit, since
a cut-off of the luminosity function due to the age of the Galaxy is expected
at approximately the same bolometric magnitude (see Figure~\ref{07}).

As we explained at the end of section~\ref{sec:6}, most of the WDs at
the bright end of the WDLF can be included with Gaia whereas they could not before,
since Gaia has a lower mean lower proper motion completeness limit than previous surveys.
From the long tail of the distance distribution (top panel of Figure~\ref{08}),
we see that there are many halo WDs with $G<20$ beyond $\sim 1$~kpc,
which all have absolute bolometric magnitudes $M_\mathrm{bol} < 8$.
It is because of the inclusion of these WDs that the bright end of the WDLF
will probably be better constrained with Gaia than ever before. 

The masses and cooling times of the $1.5 \cdot 10^3$ halo WDs with $G<20$ 
are plotted on top of the interpolated cooling track panels of Figure~\ref{03}.
Again, the same markers are used as in Figures~\ref{04} and \ref{08}.
For the halo WDs that are merger products, it is indicated what type of WD 
the merger product is (He WD or CO WD) by the particular panel the marker
is drawn in.
Plusses and diamonds represent single WDs and resolved binary WDs 
respectively, which all have a CO core, as can also be seen from Table~\ref{table:03}.
This is not surprizing since He WDs can only be formed through binary interaction
within the age of the universe and ONe WDs cool so fast that they pile up at the
faint end of the WDLF, which will not be covered by Gaia.

\begin{figure}[h!]
\centering	
	\resizebox{0.85\hsize}{!}{\includegraphics{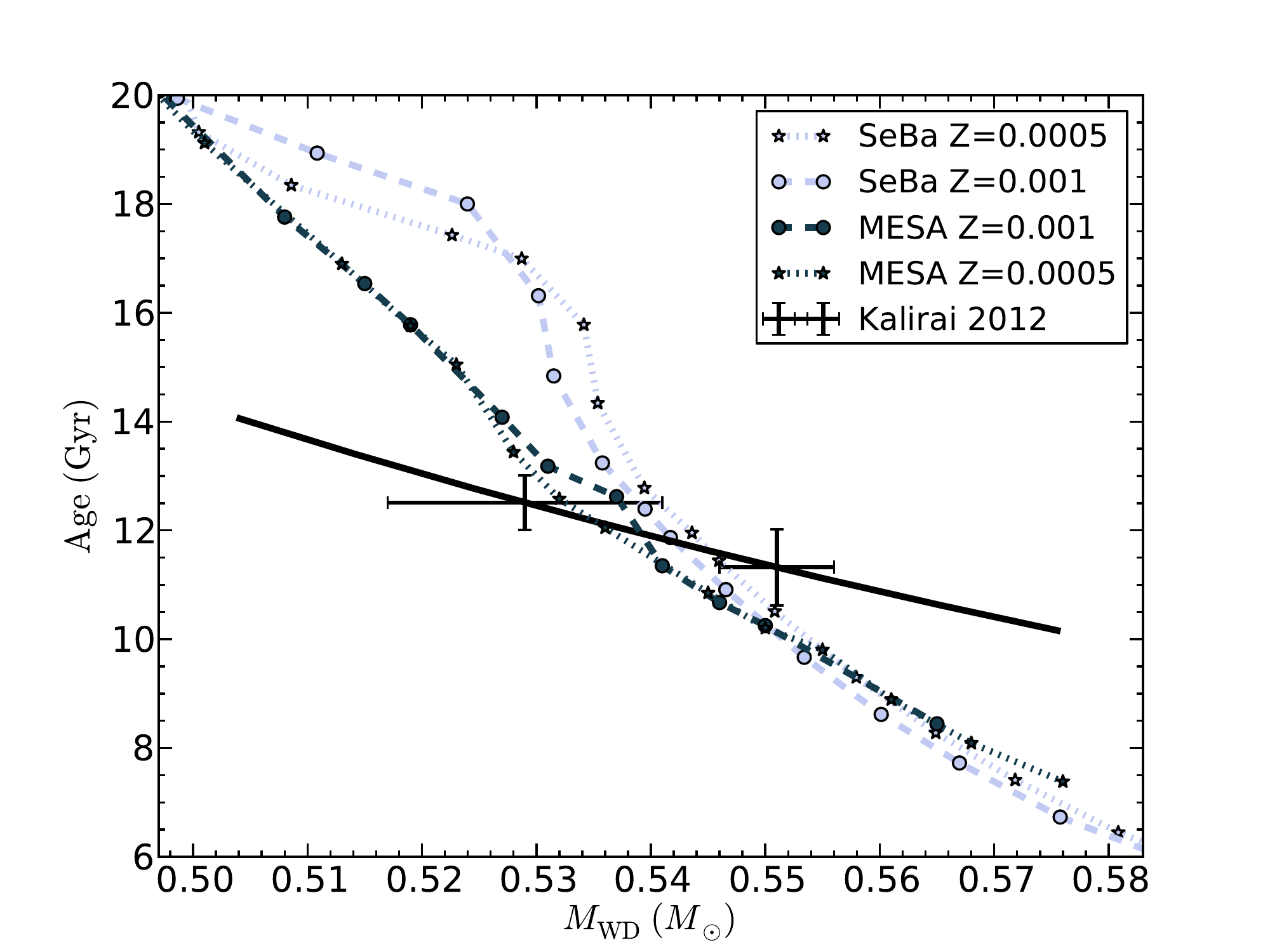}}
		\caption{Age of the brightest single halo WDs (and thus of the Galactic halo)
		as a function of WD mass. Since for single stars every age corresponds
		one-to-one to an initial stellar mass and WD mass, this is an alternative
		representation of the IFMR.	The IFMRs for single stars predicted by SeBa 
		(light blue curves) and the those predicted by MESA (dark blue curves) 
		both differ from the black solid straight line through the data points 
		with error bars of \citet{Kalirai:2012}.
		The dashed lines correspond to the metallicity value that 
		we used as our standard for the Milky Way halo
		in this study (Z=0.001), while the dotted lines correspond 
		to half that metallicity value (Z=0.0005).}
		\label{09}
\end{figure}

It is interesting to see the narrow line at $m = 0.54~\mathrm{M}_\odot$ in Figure~\ref{03},
where the single and non-interacting binary WDs pile up. This can be explained
by the evolution lifetime of single and non-interacting ZAMS stars with an initial mass
of $0.84~\mathrm{M}_\odot$, which is equal to the age of the halo in our standard model
and we see in the middle panel of Figure~\ref{05} that they become $0.54~\mathrm{M}_\odot$ WDs.
\citet[hereafter K12]{Kalirai:2012} first pointed out that the mass determination of these bright
single halo WDs can be used to determine the age of the inner halo.
The determination of the masses of the brightest WDs in a globular cluster
provides an anchor point on the IFMR for low metallicity stars, 
since their ages can be deduced independently from the cluster age. 
K12 then drew a straight line through this anchor point
and the mass of the brightest halo WDs, 
yielding a linear IFMR for WD masses between $0.50$ and $0.58~\mathrm{M}_\odot$, 
see Figure~\ref{09}. The age of the field halo stars was subsequently determined using the 
Dartmouth Stellar Evolution Database \citep{Dotter:2008},
in which $\sim 0.55~\mathrm{M}_\odot$ WDs (with $0.83~\mathrm{M}_\odot$ progenitors)
are $\sim 11.4$~Gyr old.

Although K12 did a carful analysis on the binarity of field halo WDs 
and in the globular cluster, the binary fraction in
globular clusters is generally lower than in the field
\citep[eg.][and references therein]{Ji:2013}.
Especially for the latter, the observed single WDs could thus still have 
an unseen companion or be the result of a binary merger.
Furthermore, the behaviour of the (metallicity-dependent) IFMR in this mass regime 
is not well determined yet theoretically.
This can be seen from Figure~\ref{09}, where we compare the low-metallicity IFMR of K12 with 
two IFMRs predicted by the detailed stellar evolution code MESA \citep{Paxton:2010} and with 
two IFMRs predicted by SeBa.
The relation in age and mass in SeBa shows a strong upturn in age between WD masses with 
$0.54-0.53 \ \mathrm{M}_\odot$, after which the slopes become shallower again. This difference in slopes
is due to two different evolution paths for low-mass main sequence stars. The higher mass stars follow
a standard evolution path: they become WDs after the AGB phase. The lower mass stars on the other hand,
lose their envelope on the RGB, whereafter they become WDs.
In MESA this transition between these two evolution paths is implemented differently, yielding
a more linear IFMR, which however is still steeper than the one inferred by K12. 
The observations seem to be consistent with all of these model lines.
In fact, it will be challenging to observationally distinguish between the mass-age relations predicted
by MESA and SeBa, since the difference between the two sets of lines is largest after a Hubble time.
If on the other hand, $\sim 0.51 \ \mathrm{M}_\odot$ single WDs will be found to follow the black solid line
in Figure~9, as some data seems to imply \citep[see eg. Table~1 of][]{Renzini:1996}, there is a challenge
for the theoretical modellers to explain how such low mass WDs can be formed by single stellar evolution
within the age of the universe.

This comparison between MESA and SeBa was made using AMUSE \citep{Portegies-Zwart:2009,Portegies-Zwart:2013,
Pelupessy:2013}. See also \citet{Renedo:2010} for a comparison between theoretically and observationally 
determined IFMRs with different metallicities.

\newpage
\section{Conclusions} \label{conclusions}

The easiest way to constrain the IMF with halo WDs is to determine the halo WD
number density, because the normalization of the WDLF is linked one-to-one
to the IMF. From a comparison between our derived halo WDLFs with the 
WDLF observed by RH11, we conclude that a Kroupa IMF 
($\chi^2 = 2.29$) is slightly preferred over a Salpeter IMF ($\chi^2 = 2.99$).
A top-heavy IMF ($\chi^2 = 5.74$) clearly overpredicts the number of faint halo WDs.
Due to large uncertainties on the normalization, it is not
yet possible to completely rule out a non-standard IMF.
However, also the shape of the WDLF corresponding to a top-heavy IMF 
has worse agreement with the WDLF observed by RH11 than those of the WDLFs 
corrsponding to models with a Kroupa or Salpeter IMF.
Although most investigated halo models match the observed WDLF approximately equally well
if we fix the normalization of our WDLF ($2.3 \lesssim \chi_\mathrm{min}^2 \lesssim 2.7$), none of the models
comes close to a reduced $\chi^2$ value of 1.

The exact number of halo WDs that Gaia can observe depends on how easily they
can be distinguished from thin and thick disk WDs. In our standard model we find
that Gaia will be able to detect approximately $1.5 \cdot 10^3$ halo WDs,
which is an order of magnitude more than the currently known number of halo WDs.
Taking into account selection effects will probably not reduce this number 
by more than a factor of two. A wrong assumption on the mass function of unevolved stars has
a stronger effect on the determined number density of halo WDs, but this will 
probably also be constrained by Gaia.
If our assumptions are correct, the error bars on the observationally determined
WDLF will become smaller with Gaia, at the part of the luminosity function
that already has the smallest error bars (e.g. $5 \lesssim M_\mathrm{bol} \lesssim 10$),
but also for the fainter luminosity bins. This implies that we might soon
be able to start ruling out IMFs on the basis of their predicted WD number densities,
especially in the $M_\mathrm{bol} \gtrsim 15$ luminosity bins.

Since the effect of the SFH of halo stars is the strongest
at the faint end of the WDLF, where we expect only a handful
of WDs to be detected by Gaia, it will be observationally challenging to put strong
constraints on this parameter in the near future.
Although the differences are small, from the current observational constraints,
we find that a model in which there was a burst of SF 13 Gyr ago
($\chi^2 = 2.29$, $\chi_\mathrm{min}^2 = 2.26$) is slightly preferred over 
a burst of starformation 10 Gyr ago ($\chi^2 = 2.35$, $\chi_\mathrm{min}^2 = 2.43$) and over
continuous star formation $10 - 13$ Gyrs ago 
($\chi^2 = 2.31$, $\chi_\mathrm{min}^2 = 2.37$).

A determination of the masses of the brightest halo WDs can in theory 
be used to determine the age of the halo as suggested by K12. However, at this point only
preliminary conclusions can be drawn since the observational uncertainties are large
and the effect of binarity and/or metallicity can be underestimated. It would be useful 
to have more anchor points on the low-metallicity IFMR than the current single one.

With Gaia it will be possible to constrain for the first time the bright part of the 
field halo WDLF, where contributions from (unresolved) binary WDs are considerable
\citep[that this can be done in star clusters,
and for singly evolved WDs was shown by][]{Goldsbury:2012}.
By determining the periods of WDs with masses below $\approx 0.5 \ \mathrm{M}_\odot$ 
(which can safely be assumed to be in binaries or to be the result of a binary merger)
with Gaia follow-up observations, we can start to explore how binary stars 
with low metallicities evolve.
In this paper, we do not vary any binary evolution parameters, but deviations 
from the predictions we made in the bottom panel of Figure~\ref{04} 
are produced by different models of binary evolution.

It might be possible to put some constraints on the binary fraction by the
number of pre-WDs that will be observed, although we expect this number to be small,
since a large fraction of the pre-WD candidates will be indistinguishable from main
sequence stars or giants. Furthermore, with the Bergeron models 
for WD cooling, pre-WDs are not expected to exist at all.
However, pre-WDs can help us to constrain WD cooling
models, because they are situated on an uncertain part of the WD cooling track, 
in the early phases of WD cooling. 

If future observations on halo WDs go up to fainter magnitudes than Gaia can observe, 
we will be able to determine the validity of a top-heavy IMF or the Bergeron cooling models.
In this respect, observations of the Large Synoptic Survey Telescope (LSST) will be very helpful
\citep[over $4 \cdot 10^5$ halo WDs to $r < 24.5$;][]{LSST-Science-Collaboration:2009}.
To improve this study, WD cooling tracks and corresponding colours and magnitudes over the
whole parameter range of WD masses and cooling times would be useful, for WDs
with low-metallicity progenitors. 
In the near future, we will couple a semi-analytic model for galaxy formation
with a binary population synthesis code and study how this affects the halo WD population.

\begin{acknowledgements} 
We thank Else Starkenburg as well as
the anonymous referee for valuable comments.
We thank the Netherlands Research School for Astronomy
(NOVA) and the Netherlands Organisation for Scientific Research 
(NWO) for financial support (\#639.073.803 [VICI] and
\#614.061.608 [AMUSE]).
AH acknowledges financial support from ERC StG 240271, Galactica.
\end{acknowledgements}

\bibliographystyle{aa}
\bibliography{example}

\appendix
\section{} \label{Ap:A}

Cartesian coordinates are related to spherical coordinates by 
\begin{eqnarray}
&x &= r \sin \theta \cos \phi  \\ 
&y &= r \sin \theta \sin \phi  \\ 
&z &= r \cos \theta 
\end{eqnarray}
with radius $r$, polar angle $\theta$ and azimuth angle $\phi$.
The Sun is assumed to be at position $(x,y,z)_\odot = (r_0,0,0)$, or
equivalently at $(r, \theta, \phi)_\odot = (r_0,\pi/2,0)$.
We define the primed coordinates
\begin{eqnarray}
&x' &= r' \sin \theta' \cos \phi \equiv x \\ 
&y' &= r' \sin \theta' \sin \phi \equiv y \\ 
&z' &= r' \cos \theta' \equiv q \ z
\end{eqnarray}
such that the local halo density (equation~(\ref{eq.1})) can be expressed independent of 
a polar angle and azimuth angle:
\begin{equation}
\rho(x',y',z') \equiv \rho(r') = \rho_0 \left(\frac{r'}{r_0}\right)^n.
\end{equation}
We note that in the Galactic plane, $z=0$, thus primed radius $r'=r$ and 
the primed polar angle $\theta' = \theta = \nicefrac{\pi}{2}$.
At the Galactic pole, $\theta' = \theta = 0$, and $r' = z' = q \ z = q \ r$.
In all other cases, the relation between the $r'$,  $\theta'$ and their spherical equivalents
is given by 
\begin{eqnarray}
&\theta' &= \arctan\left(\frac{\tan(\theta)}{q}\right)  \label{theta'}\\ 
&r' &= r \ \frac{\sin \theta}{\sin \theta'}. \label{r'}
\end{eqnarray}

Since we assume an oblate stellar halo ($q <1$), it follows from equations \ref{theta'} and \ref{r'} 
that $\theta' \ge \theta$ and $r'\le r$ for any given point in the spheroid.
Because we want a sphere with radius $\xi$ around the Sun to be contained in our simulated area, 
we set the boundary conditions,
\begin{eqnarray}
r_0-\xi &\le r &\le r_0+\xi  \\ 
\delta &\le \theta &\le \nicefrac{\pi}{2} - \delta \\
-\epsilon &\le \phi &\le \epsilon, 
\end{eqnarray}
with $\delta \le \arctan(\nicefrac{r_0 q}{\xi})$ and $\epsilon \le \arctan(\nicefrac{\xi}{r_0})$.
These set the limits of integration in our determination of the stellar halo mass:
\begin{eqnarray}
M = \frac{\rho_0}{r_0^n}\int_{r_0-\xi}^{r_0+\xi} r^{n+2} \mathrm{d}r 
\int_{\delta}^{\nicefrac{\pi}{2}-\delta}\left(\frac{\cos^2\theta}{q^2}+\sin^2\theta\right)^{\nicefrac{n}{2}} \sin \theta \ \mathrm{d}\theta 
\int_{-\epsilon}^{\epsilon} \mathrm{d}\phi. \nonumber \\
\end{eqnarray}
In order to solve the integral over $\theta$, we now first make an estimation of $\delta$.
With the assumed values of $\xi$, $q$ and $r_0$ mentioned in the main text,
we find $\delta \le 0.334 \ \pi$. Thus, we take $\delta = \pi/3$.
The integral over $\theta$ can now be expressed as the hypergeometric function
${}_2F_1\left(\nicefrac{1}{2},-\nicefrac{n}{2};\nicefrac{3}{2};\nicefrac{1}{4}-\nicefrac{1}{4q^2}\right)$.
Again with $q = 0.64$ and $n=-2.8$ for consistency with \citet{Juric:2008},
we find ${}_2F_1\left(0.5,1.4;1.5;-0.36\right) = 0.866$. 
Because this value of $n\ne -3$, the integral over $r$ can also be evaluated:
\begin{eqnarray}
\frac{1}{r_0^n}\int_{r_0-\xi}^{r_0+\xi} r^{n+2} \mathrm{d}r = \frac{(r_0+\xi)^{n+3} - (r_0-\xi)^{n+3}}{(n+3) \ r_0^n} = 3.91 \cdot 10^{11}\ \mathrm{pc}^{3}. \nonumber \\
\end{eqnarray}
The integral over $\phi$ yields $2\epsilon$, thus after choosing $\epsilon = \arctan(\nicefrac{\xi}{r_0})$ 
this reads $2\arctan(\nicefrac{\xi}{r_0}) = 0.707$.
The multiplication of an assumed value of $\rho_0 = 1.5 \cdot 10^{-4} \ \mathrm{M}_\odot \ \mathrm{pc}^{-3}$ \citep{Fuchs:1998}
with these three integrals gives $M_\mathrm{unev} = 3.6 \cdot 10^7 \ \mathrm{M}_\odot$.

\section{} \label{Ap:B}

In case $\phi(m)$ is a single power law function between the upper and lower mass boundary of unevolved stars
in our simulation box $m_\mathrm{high,unev}$ and $m_\mathrm{low,unev}$, the total mass in unevolved stars
\begin{equation}
 \frac{M_\mathrm{unev}}{\mathrm{M}_\odot} = \int_{m_\mathrm{low,unev}}^{m_\mathrm{high,unev}} A \ m^{-\gamma_\mathrm{unev}} \mathrm{d}m 
= \frac{A\left( m_\mathrm{high,unev}^{1-\gamma_\mathrm{unev}}-m_\mathrm{low,unev}^{1-\gamma_\mathrm{unev}} \right)}{1-\gamma_\mathrm{unev}}. \label{B1}
\end{equation}
Given the mass in unevoloved stars $M_\mathrm{unev}$ which was derived in Appendix~\ref{Ap:A},
$\gamma_\mathrm{unev} = -1$, $m_\mathrm{high,unev} = 0.8$ and $m_\mathrm{low,unev} = 0.1$, 
this results in a normalization constant belonging to the lower limit on the number of unevolved (single) stars $N_\mathrm{unev}$
in our simulation box
$A_\mathrm{lower} = 1.1 \cdot 10^8$.
When substituted into equation~(\ref{N}), this yields
\begin{equation}
N_\mathrm{unev} > A_\mathrm{lower} \left( m_\mathrm{high,unev} - m_\mathrm{low,unev} \right) = 8.0 \cdot 10^7.
\end{equation}

We derive an upper limit on the number of evolved stars $N_\mathrm{ev}$ in our simulation box,
for the three different IMFs that we investigate in this paper by determining their normalization constants
from the IMF at $m_\mathrm{high,unev}$.
For example, writing the normalization constant for the upper limit on the number of evolved stars in case of a
Kroupa IMF as $B_\mathrm{upper}$, the relation $\phi(m_\mathrm{high,unev}) = A_\mathrm{lower} = B_\mathrm{upper} \ (m_\mathrm{high,unev})^{-2.2}$ 
leads to $B_\mathrm{upper}=7.0 \cdot 10^7$,
from which follows
\begin{equation}
N_\mathrm{ev, Kroupa} < B_\mathrm{upper} \cdot I_\mathrm{ev, Kroupa} = 5.9 \cdot 10^7,
\end{equation}
where 
\begin{equation}
I_\mathrm{ev, Kroupa} = \int_{0.8}^{1.0} m^{-2.2} \mathrm{d}m + \int_{1.0}^{100} m^{-2.7} \mathrm{d}m = 0.84.
\end{equation}
To obtain actual numbers instead of an upper limit, we assume that
the low-mass part of the IMF is correctly given by equation~(\ref{Kroupa}), with
normalization constant $B$,
\begin{equation}
\frac{M_\mathrm{unev}}{\mathrm{M}_\odot} = \int_{0.1}^{0.5} \frac{35}{19} \ B \ m^{-0.3} \mathrm{d}m + \int_{0.5}^{0.8} B \ m^{-1.2} \mathrm{d}m = 1.6 \ B,
\end{equation}
again using the calculated total mass in unevolved stars $M_\mathrm{unev} = 3.6 \cdot 10^7 \ \mathrm{M}_\odot$,
we find $B = 2.2 \cdot 10^7$. Now because 
\begin{equation}
I_\mathrm{unev, Kroupa} = \int_{0.1}^{0.5} \frac{35}{19} \ m^{-1.3} \mathrm{d}m + \int_{0.5}^{0.8} m^{-2.2} \mathrm{d}m = 5.5
\end{equation}
we find
\begin{eqnarray}
&N_\mathrm{unev, Kroupa} &= B \cdot I_\mathrm{unev, Kroupa} = 1.2 \cdot 10^8 \\
&N_\mathrm{ev, Kroupa} &= B \cdot I_\mathrm{ev, Kroupa} = 1.9 \cdot 10^7.
\end{eqnarray}

Assuming that the Salpeter IMF holds for masses $m > 0.8$
results in the same way into an upper limit on the number of evolved stars,
whereas assuming that it is for the entire mass range $0.1 < m < 100$ gives
the expected number of evolved stars.
Since
\begin{equation}
I_\mathrm{ev, Salpeter} = \int_{0.8}^{100} m^{-2.35} \mathrm{d}m = 1.00,
\end{equation}
the upper limit on the number of evolved stars in the case of a
Salpeter IMF immediately follows from the normalization constant
$C_\mathrm{upper} = A_\mathrm{lower}/m_\mathrm{high,unev}^{-2.35} = 6.7 \cdot 10^7$,
\begin{equation}
N_\mathrm{ev, Salpeter} < C_\mathrm{upper} \cdot I_\mathrm{ev, Salpeter} = 6.7 \cdot 10^7.
\end{equation}
The expected number of stars in our simulation box if the low-mass part of the mass function is also Salpeter
\begin{eqnarray}
&N_\mathrm{unev, Salpeter} &= C \cdot I_\mathrm{unev, Salpeter} = 1.7 \cdot 10^8 \\
&N_\mathrm{ev, Salpeter} &= C \cdot I_\mathrm{ev, Salpeter} = 1.1 \cdot 10^7
\end{eqnarray}
with
\begin{equation}
\frac{M_\mathrm{unev}}{\mathrm{M}_\odot} = \int_{0.1}^{0.8} C \ m^{-1.35} \mathrm{d}m = 3.3 \ C,
\end{equation}
thus $C= 1.1  \cdot 10^7$, and
\begin{equation}
I_\mathrm{unev, Salpeter} = \int_{0.1}^{0.8} m^{-2.35} \mathrm{d}m = 15.6.
\end{equation}

Finally, for the top-heay IMF we derive the normalization constants for
the Komiya IMF (indicated by the letter $D$) and the Salpeter IMF (indicated by the letter $E$)
simultaneously, using the MDF of the halo described by \citet{An:2013}, who studied halo main-sequence stars with 
masses between 0.65 M$_\odot$ and 0.75 M$_\odot$ in the Sloan Digital Sky Survey.
These authors found that the halo can be described by a two-component model,
with 24\% of the stars belonging to a low-metallicity population with a peak at [Fe/H]$ \ = -2.33$
(i.e. their calibration model).
If this population of low-metallicity stars is born according to a Komiya IMF, 
we have
\begin{equation}
\int_{0.65}^{0.75} D \exp\left[-\frac{\log_{10}^2 (m/\mu)}{2\sigma^2} \right] \frac{\mathrm{d} m}{m} = \frac{0.24}{0.76} \int_{0.65}^{0.75} E m^{-2.35} \mathrm{d}m,\label{B18}
\end{equation}
which holds for and $D$ and $E$, as well as for $D_\mathrm{upper}$ and $E_\mathrm{upper}$. 
The normalization constants for the upper limit on the number of evolved stars in case of a
top-heavy IMF follow again from
\begin{eqnarray}
&\phi(m_\mathrm{high,unev}) &= \frac{D_\mathrm{upper}}{m_\mathrm{high,unev}} \ \exp\left[-\frac{\log_{10}^2 (m_\mathrm{high,unev}/\mu)}{2\sigma^2} \right] \nonumber \\
&& + \ E_\mathrm{upper} \ (m_\mathrm{high,unev})^{-2.35} = A_\mathrm{lower}.
\end{eqnarray}
From the standard integral
\begin{equation}
\int \exp\left[-\frac{\log_{10}^2 (m/\mu)}{2\sigma^2} \right] \frac{\mathrm{d}m}{m} = \frac{\sqrt{\pi/2} \ \sigma \ \mathrm{erf} \left[\nicefrac{\log_{10} (m/\mu)}{\sqrt{2}\sigma}\right]}{\log_{10} e}
\end{equation}
it now follows that $D_\mathrm{upper} = 1.4 \cdot 10^9$ and $E_\mathrm{upper} = 4.3 \cdot 10^7$.
Consequently, the number of evolved stars
\begin{eqnarray}
&&N_\mathrm{ev, Komiya} < D_\mathrm{upper}  \cdot I_\mathrm{ev, Komiya} = 3.3 \cdot 10^9 \\
&&N_\mathrm{ev, Salpeter \ (top-heavy)} < E_\mathrm{upper} \cdot I_\mathrm{ev, Salpeter} = 4.3 \cdot 10^7
\end{eqnarray}
with 
\begin{equation}
I_\mathrm{ev, Komiya}  = \int_{0.8}^{100} \exp\left[-\frac{\log_{10}^2 (m/\mu)}{2\sigma^2} \right] \ \frac{\mathrm{d} m}{m} = 2.29.
\end{equation}
If the suggested top-heavy IMF holds in the low-mass regime,
\begin{eqnarray}
 \frac{M_\mathrm{unev}}{\mathrm{M}_\odot} &&= \int_{0.1}^{0.8} D \exp\left[-\frac{\log_{10}^2 (m/\mu)}{2\sigma^2} \right] \mathrm{d}m + \int_{0.1}^{0.8} E m^{-1.35} \mathrm{d}m \nonumber \\
&& = 4.4 \cdot 10^{-3} D + 3.3 E, \label{B25}
\end{eqnarray}
where we used the standard integral: $\int \exp\left[-\frac{\log_{10}^2 (m/\mu)}{2\sigma^2} \right] \ \mathrm{d} m =$
\begin{equation}
- \frac{\sqrt{\pi/2} \ \mu \sigma \exp\left(\frac{\sigma^2}{2\log_{10}^2 e} \right)}{\log_{10} e} \ \mathrm{erf} \left[\frac{\sigma^2 - \log_{10} e \log_{10} (m/\mu)}{\sqrt{2}\sigma \log_{10} e} \right].
\end{equation}
Combining equations \ref{B18} and \ref{B25}, we find
$D = 3.4 \cdot 10^8$ and $E = 1.0 \cdot 10^7$, as well as
\begin{eqnarray}
&&N_\mathrm{unev, Komiya} = D \cdot I_\mathrm{unev, Komiya} = 2.4 \cdot 10^6 \\
&&N_\mathrm{unev, Salpeter \ (top-heavy)} = E \cdot I_\mathrm{unev, Salpeter} = 1.6 \cdot 10^8 \\
&&N_\mathrm{ev, Komiya} = D \cdot I_\mathrm{ev, Komiya} = 7.9 \cdot 10^8 \\
&&N_\mathrm{ev, Salpeter \ (top-heavy)} = E \cdot I_\mathrm{ev, Salpeter} = 1.0 \cdot 10^7,
\end{eqnarray}
where
\begin{equation}
I_\mathrm{unev, Komiya}  = \int_{0.1}^{0.8} \exp\left[-\frac{\log_{10}^2 (m/\mu)}{2\sigma^2} \right] \ \frac{\mathrm{d} m}{m} = 7.0 \cdot 10^{-3}.
\end{equation}

\end{document}